\documentclass[reprint,amsmath,amssymb,aps,prx,tightenlines,nobibnotes,superscriptaddress]{revtex4-2}

\usepackage{graphicx}
\usepackage{float}
\usepackage{dcolumn}
\usepackage{bm}
\usepackage{xcolor}
\usepackage{natbib}
\usepackage{setspace}
\usepackage{multirow}
\usepackage{amsmath}              % Mathematikbefehle
\usepackage{amssymb}              % Unter anderem schöne Mengensymbole
\usepackage{amsfonts}             % Weitere Symbole für equation Umgebung
\usepackage[amssymb]{SIunits}     % Einheiten per Befehl einfügen
\usepackage[applemac]{inputenc}
\usepackage[english]{babel}
\newcommand{\edp}{{\mathbf{e}_{\phi}}}

\newcommand{\eu}{{\bar{\mathbf{e}}_u}}
\newcommand{\ep}{{\bar{\mathbf{e}}_{\phi}}}
\newcommand{\TT}{\mathbf{T}}
\newcommand{\ebar}{{\bar{\mathbf{e}}}}
\renewcommand{\div}{\mathrm{div}}

\newcommand{\bteun}{(\bar{T}_e)_u^n}
\newcommand{\bteuu}{(\bar{T}_e)_{uu}}
\newcommand{\btepp}{(\bar{T}_e)_{\phi \phi}}
\newcommand{\btduu}{(\bar{T}_d)_{uu}}
\newcommand{\btdpp}{(\bar{T}_d)_{\phi \phi}}
\newcommand{\btdup}{(\bar{T}_d)_{u \phi}}
\newcommand{\Cuu}{\frac{\psi'}{h}}
\newcommand{\Cpp}{\frac{\sin\psi}{r}}
\newcommand{\bvuu}{\bar{S}_{uu}}
\newcommand{\bvpp}{\bar{S}_{\phi \phi}}
\newcommand{\bvup}{\bar{S}_{u\phi}}
\newcommand{\fn}{f^{\text{ext}}_n}
\newcommand{\fu}{f^{\text{ext}}_u}
\newcommand{\fp}{f^{\text{ext}}_{\phi}}
\newcommand{\bS}{\mathbf{S}}
\newcommand{\tbS}{\tilde{\mathbf{S}}}
\newcommand{\isoS}{\mathbf{S}^{\mathrm{iso}}}
\newcommand{\bvu}{\bar{v}_u}
\newcommand{\bvp}{\bar{v}_{\phi}}

\newcommand{\fk}{f_{\kappa}}

\begin{document}
\title{Self-organized dynamics and emergent shape spaces of active isotropic fluid surfaces}

\author{Da Gao}
\affiliation{Department of Physics, College of Physical Science and Technology, Xiamen University, Xiamen 361005, People's Republic of China}
\author{Huayang Sun}
\affiliation{Department of Physics, College of Physical Science and Technology, Xiamen University, Xiamen 361005, People's Republic of China}
\author{Rui Ma}
\email{ruima@xmu.edu.cn}
\affiliation{Department of Physics, College of Physical Science and Technology, Xiamen University, Xiamen 361005, People's Republic of China}
\author{Alexander Mietke}
\email{alexander.mietke@physics.ox.ac.uk}
\affiliation{Rudolf Peierls Centre for Theoretical Physics, Department of Physics, University of Oxford, Parks Road, Oxford OX1 3PU, United Kingdom}

\begin{abstract}
Theories of self-organized active fluid surfaces have emerged as an important class of minimal models for the shape dynamics of biological membranes, cells and tissues. However, due to their inherent geometric nonlinearities and the absence of general minimization principles in active systems, it remains a major challenge to systematically study the emergent shape spaces such theories give rise to. Here, we introduce a novel variational approach that allows for a direct computation of stationary surface geometries and flows, which enables the classification of non-equilibrium phase transitions in shape spaces described by active surface theories. To achieve this, we construct a dissipation functional systematically from the entropy production in active surfaces and show how generic symmetries imposed by Onsager relations can be exploited to also account for reactive non-dissipative terms in constitutive laws. This functional is supplemented by Lagrange multipliers that relax nonlinear geometric constraints, which leads to a tractable variational problem suitable for implicit dynamic simulations and explicit calculations of non-trivial steady state geometries and flows. We apply this framework to study the dynamics of open fluid membranes and closed active fluid surfaces, and characterize the space of stationary solutions that corresponding surfaces and flows occupy. These analyses rationalize the interplay of first-order shape transitions of internally and externally forced fluid membranes, reveal degenerate regions in stationary shape spaces of mechanochemically active surfaces and identify a mechanism by which hydrodynamic screening controls the geometry of active surfaces undergoing cell division-like shape transformations. 

\end{abstract}

\keywords{Surface flows \and Active surfaces \and Cell division}

\maketitle
\newpage

\setlength{\parindent}{0pt}

\section{Introduction}
Many crucial processes in living systems rely on the geometric control of two-dimensional structures with fluid-like properties. Membranes, for example, are made of thin lipid bilayer sheets and are critical in regulating traffic into and out of cells via endocytosis and exocytosis~\hbox{\cite{kaksonen2018mechanisms,jahn1999membrane,wu2014exocytosis}}. Due to weak hydrophobic interactions~\cite{Albert2002book,phillips2012physical}, these lipid sheets behave as two-dimensional fluids~\cite{Dimova_2006}, while orientational order in lipid packing allows membranes to resist bending. The cortex of eukaryotic cells is a thin polymer filament network beneath the cell membrane~\cite{Albert2002book} that can also behave as an effectively two-dimensional viscous fluid due to filament and cross-linker turnover on time scales of seconds~\cite{salbreux2012actin}. Active mechanical forces generated by motor proteins lead to cortical flows~\cite{bray1988cortical,mayer2010anisotropies}, and play a crucial role in maintaining cell shape and enabling dynamic cellular processes such as migration and division~\cite{lecuit2007cell,salbreux2012actin,Paluch2017NCB,mayer2010anisotropies,Turlier2014}. Finally, epithelial tissues are effectively two-dimensional cell monolayers that are critical for embryonic development and morphogenesis~\cite{lecuit2007cell,martin2010integration}. This role is facilitated by active cellular processes that allow epithelial tissues to autonomously change their shape~\cite{heisenberg2013forces}. Cellular rearrangements, cell division and cell death lead to viscous material properties on the time scale of hours~\cite{ranft10}. The ability to resist bending plays an important role in both the cellular cortex~\cite{dead23} and in epithelial tissues~\cite{denk21}. 

Given the abundance of fluid surfaces involved in the shape-regulation of cells and tissues, minimal models of self-organizing active fluid surfaces have become an essential framework to understand the emergence of shape in living systems~\cite{Guillaume2017,Mietke2019Self,Mietke2019Minimal,Salbreux2022,Salbreux2023elife,bail25}. An often encountered bottleneck in investigating such theories is the inevitably nonlinear nature of surface force balance equations that involve \textit{a priori} unknown, or even dynamic, domain geometries. Restricting oneself to axisymmetric geometries and surfaces that exhibit a purely passive response to bending, this bottleneck has been addressed within a historical body of work starting from the formulation of the Helfrich bending free energy~\cite{canham1970minimum,Helfrich1973,evans1974bending} over the derivation of highly non-linear shape equations~\cite{ouyang1987,OuYang1989} to the formulation of a variational approach that converts the shape equation into a tractable boundary value problem~\hbox{\cite{lipowsky1991conformation,seif91,ouyang1993,julicher1994}}. The success of this variational formulation in explaining many experimental observations~\cite{lipowsky1993domain,julicher1996,julicher2002,seifert1997configurations,hassinger2017design, walani2015endocytic,chernomordik2008mechanics} arose from its ability to systematically characterize the space of stationary surface shapes a given membrane theory gives rise to. 

Active materials can in general not be described in terms of an energy functional, which raises the question if and how variational formulations -- analog to those previously developed for passive membranes -- can be systematically developed for out-of-equilibrium fluid surfaces. A promising route to address this question has been pioneered in~\cite{Arroyo2009PRE,Doi_2011,Arroyo2018} using Onsager's variational principle. In this formulation, a Rayleigh functional is constructed that includes the free energy change rate and dissipative energy losses. Variation of the Rayleigh functional then gives rise to a set of dynamic equations~\cite{Arroyo2018} that should be equivalent to force and moment balance equations in terms of some constitutive laws. However, while suitable expressions for common contributions to dissipation, such as viscous shear stresses, can essentially be guessed, there exists currently no systematic way to start from more general constitutive laws of active surfaces~\cite{Guillaume2017,GuillaumePRR2022} and formulate a corresponding functional. Additionally, because the minimization of dissipation is not a fundamental principle~\cite{katch65}, we have to ask if such functionals exist at all in more general cases. 

Here, we build on recent theoretical advances in applying the framework of irreversible thermodynamics to curved and deforming surfaces~\cite{Guillaume2017,GuillaumePRR2022} to address these questions. Exploiting the symmetries imposed by Onsager relations~\cite{onsa31,onsa31b,Julicher_2018}, we show that the entropy production of active deforming surfaces derived in these works can indeed be used to construct suitable Rayleigh functionals. Specifically, variations of the latter give rise to a boundary value problem that is equivalent to the force and moment balance equations for the full set of \textit{a priori} defined constitutive laws, including those describing the out-of-equilibrium maintenance of chemical potentials, and even if non-dissipative active couplings are present. 

A key numerical challenge when solving for the shape dynamics of a deforming surfaces is that also its parameterization will depend on time. This potentially results in a lack of computational control over the accuracy with which the physical surface can be represented and is often addressed using arbitrary Lagrangian-Eulerian (ALE) parameterizations~\cite{Arroyo_2019,SAHU2020,Aland2023}. In this work, we design an ALE parameterization that can directly be integrated into the dissipation functional, facilitates a numerically robust implicit solution of the shape dynamics and ensures points on the deforming surface are mapped to fixed reference coordinates via a simple scaling. We will use this integrated framework to study both the dynamics and the space of stationary geometries and flows of active fluid surfaces undergoing biologically relevant shape transformations. The numerical implementations of this framework can be found on GitHub~\cite{git}.

\section{Dynamics of parameterized curved surfaces}
We describe a surface $\Omega$ by the vector field \hbox{$\mathbf{X}(s^1,s^2,t)\subset\mathbb{R}^3$} that is parameterized by two generalized coordinates \hbox{$s^i$ ($i\in\left\{1,2\right\}$}) and time~$t$. Tangent vectors and surface normal are given by $\mathbf{e}_i=\partial_i\mathbf{X}$ and $\mathbf{n}=\mathbf{e}_1\times\mathbf{e}_2/|\mathbf{e}_1\times\mathbf{e}_2|$, respectively, where we denote $\partial_i:=\partial/\partial s^i$. Metric tensor and curvature tensors are defined by $g_{ij}=\mathbf{e}_i\cdot\mathbf{e}_j$ and $C_{ij}=-\mathbf{n}\cdot \partial_i\partial_j \mathbf{X}$. The surface area element is given by $dA=\sqrt{g}ds^1ds^2$, where $g=\det(g_{ij})$. Vector fields on the surface $\mathbf{w}=\mathbf{w}_{\parallel}+\mathbf{w}_{\perp}$ can be expanded in terms of their in-plane components $\mathbf{w}_{\parallel}=w^i\mathbf{e}_i$ and their normal components $\mathbf{w}_{\perp}=w_n\mathbf{n}$. We denote Euclidean representations of metric and curvature tensor by $\mathbf{G} = g_{ij}\mathbf{e}^i \otimes \mathbf{e}^j$ and $\mathbf{C} = C_{ij}\mathbf{e}^i \otimes \mathbf{e}^j$, respectively, and similarly for other tangential tensors. 

\subsection{Parameterization of deforming surfaces}
Any motion and deformation of the surface can be described by a local center-of-mass velocity field $\mathbf{v}=v^i\mathbf{e}_i+v_n\mathbf{n}$, where the components ($v^i$) and ($v_n$) correspond to in-lane flows surface deformations, respectively. The general surface dynamics is given~by
\begin{equation}\label{eq:GenDyn}
\frac{d\mathbf{X}}{dt}=\mathbf{v},
\end{equation}
where $\frac{d}{dt}(\cdot)$ denotes the total time derivative. To discuss the appearance of a time-dependent parameterization in Eq.~(\ref{eq:GenDyn}) explicitly, we expand the total time derivative as
\begin{equation}\label{eq:ExpGenDyn}
\partial_t\mathbf{X}+q^i\partial_i\mathbf{X}=\mathbf{v}.  
\end{equation}
The components $q^i$ of the coordinate flow $\mathbf{q}=q^i\mathbf{e}_i$ capture the in-plane motion of a surface point $\mathbf{X}$ at fixed coordinates~$s^i$, relative to its actual in-plane center-of-mass velocity $\mathbf{v}_{\parallel}=v^i\mathbf{e}_i$. For example, for a fixed material coordinate~$S$, a surface point~$\mathbf{X}$ moves by definition already with in-plane center-of-mass velocity~$\mathbf{v}_{\parallel}$, which implies $q^i=0$. This corresponds to a Lagrangian parametrization for which the surface dynamics Eq.~(\ref{eq:GenDyn}) becomes \hbox{$\partial_t\mathbf{X}=\mathbf{v}$}~\cite{Turlier2014,Guillaume2017,Salbreux2023elife}. Another common parametrization choice is to set $q^i=v^i$, corresponding to an Eulerian parametrization for which the surface dynamics Eq.~(\ref{eq:GenDyn}) becomes $\partial_t\mathbf{X}=v_n\mathbf{n}$~\cite{Mietke2019Self,Guillaume2017}. In both cases, it is the surface flow that dictates the dynamics of the parameterization. In general, however, either of the two -- $q^i(s^i,t)$ or $s^i(S,t)$ -- can be chosen freely and are related by
\begin{equation}\label{eq:qidyn}
q^i=\left.\frac{\partial s^i}{\partial t}\right|_{S}.
\end{equation}
The parameterizations resulting from different choices will typically be neither Lagrangian nor Eulerian, but as such may be fully decoupled from flows and deformations. Analogue approaches in "flat" Euclidean space are called arbitrary Lagrangian-Eulerian (ALE) methods~\cite{hirt74}, a terminology that we will adopt in the present work when extending it to embedded curved surfaces.

%  Formally, defining how the parameterization of a deforming surface evolves over time will define~$q^i$ and \textit{vice versa}.
%%%%%%%%%%%%%%%%%%
%%%%%%%%%%%%%%%%%%
\begin{figure*}[t]
	\includegraphics[width = 2.05\columnwidth]{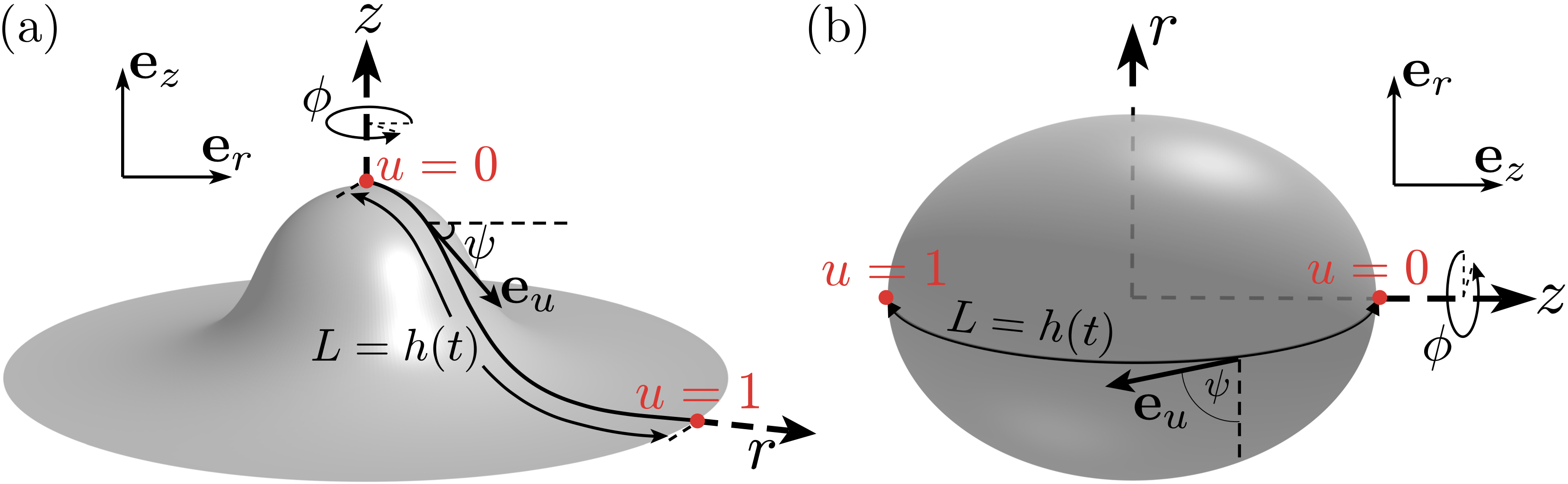}\vspace{-0.1cm}
	\caption{Parameterization of open (a) and closed (b) axisymmetric surfaces $\mathbf{X}(u,\phi,t)$ considered in this work [see Eq.~(\ref{eq:Xaxi})]. We introduce a scaling Lagrangian-Eulerian parameterization (SLE, Sec.~{\ref{sec:SLE}}) that maps fixed mesh coordinates $u\in[0,1]$ to physical arc length coordinates $s$ on the deforming surface via $s=h(t)u$. $\psi$ denotes the tangent angle and $\mathbf{e}_u=\partial_u\mathbf{X}$ the meridional tangent vector. Cylindrical coordinates $(r,\phi,z)$ parameterize the embedding Euclidean space. The total length $L=h(t)$ of the meridional outline changes in time as surfaces~deform.}
	\label{fig:SurfSketch}
\end{figure*}
%%%%%%%%%%%%%%%%%%
%%%%%%%%%%%%%%%%%%

\subsection{Conservation laws on ALE-parameterized curved surfaces}
To formulate conservation laws and transport equations on ALE-parameterized curved surfaces, one has to determine the dynamics of area integrals of the form
\begin{equation}\label{eq:Forig}
F(t)=\int_{\omega(t)}ds^1ds^2\sqrt{g}\,f(s^1,s^2,t),
\end{equation}
where the integration domain $\omega(t)$ is some coordinate region that parameterizes a fixed set of material elements. Importantly, $\omega(t)$ is only independent of time if Lagrangian material coordinates are used. We show in App.~\ref{app:movbound} that for an ALE-parameterized curved surface
\begin{equation}\label{eq:dtFfin2main}
\frac{dF}{dt}=\int_{\Omega}dA\left[\frac{1}{\sqrt{g}}\partial_t(f\sqrt{g})+\nabla_i(fq^i)\right],
\end{equation}
where $\nabla_i$ denotes the covariant derivative. Using the covariant Stokes theorem~\cite{Guillaume2017}, the second term can be identified as a boundary term that results from the dynamics of the coordinate region $\omega(t)$. As expected, this boundary contribution vanishes for a Lagrangian parameterization ($q^i=0$) and it can be interpreted as a flux $j^i_f=fv^i$ across the boundary for a Eulerian parameterization ($\smash{q^i=v^i}$). In general, however, it is not a physical flux but an artifact of any parameterization that is decoupled from the local center-of-mass motion.

As an important example, we consider $f=\rho$, where $\rho$ is the mass density of the surface, we have the total mass $F(t)=M(t)$ and mass conservation directly follows from Eq.~(\ref{eq:dtFfin2main})~(App.~\ref{app:masscons}). If multiple chemical species $\alpha$ of concentration $c_{\alpha}(s^1,s^2,t)$ are present, this leads to continuity equations for each species of the form 
\begin{equation}
    \label{eq:particle}
    \partial_tc_{\alpha}+q^i\partial_ic_{\alpha}+c_{\alpha}\div(\mathbf{v}) + \div(\mathbf{j}_{\alpha}) = r_{\alpha} + J_{n,\alpha},
\end{equation}
where $\div\left(\mathbf{\cdot}\right):=\mathbf{e}^i\cdot\partial_i\left(\mathbf{\cdot}\right)$ denotes the divergence operator on the surface, implying
\begin{equation}\label{eq:divv}
\div(\mathbf{v})=\nabla_iv^i+C_i^{\,i}v_n,   
\end{equation}
$\mathbf{j}_{\alpha}=j_{\alpha}^i\mathbf{e}_i$ are diffusive fluxes relative to the in-plane center-of-mass flux $c_{\alpha}v^i$, $r_{\alpha}$ denotes chemical reaction rates among species, and $J_{n,\alpha}$ denotes fluxes between the surface and its environment. 

\subsection{Deforming axisymmetric surfaces}
Purely Lagrangian and Eulerian parameterizations relate the coordinate flow $q^i$ only to the in-plane center-of-mass velocity~$\mathbf{v}_{\parallel}$ and are consequently slave to the systems dynamics, precluding any explicit level of control whenever surface movements and deformations occur. When solving Eq.~(\ref{eq:GenDyn}) numerically, this lack of control over the parameterization is typically countered by adaptive remeshing~\cite{Turlier2014,Aland2023}, which repeatedly interpolates the surface state and effectively resets the parameterization. By integrating surface deformations into the coordinate flow~$q^i$, we develop and apply in the following an approach that reduces the dependence of the parameterization on the surface dynamics to a global scaling, which naturally eliminates the need for flow-dependent remeshing. To make this concrete, we consider a deforming axisymmetric surface described~by
\begin{equation}
\label{eq:Xaxi}
\mathbf{X}(u, \phi, t)=[r(u,t)\cos\phi,r(u,t)\sin\phi, z(u, t)],
\end{equation}
where $r(u,t)$ and $z(u, t)$ denote time-dependent components of points on the surface, and $u \in [0, 1]$ and $\phi\in [0, 2\pi]$ are time-independent coordinates we refer to as \textit{mesh coordinates} (Fig.~\ref{fig:SurfSketch}). For general center of mass flows
\begin{equation}
\label{eq:Vel}
\mathbf{v}(u,\phi,t) = v^u(u,t) \mathbf{e}_u+v^\phi(u,t) \mathbf{e}_\phi + v_n(u,t) \mathbf{n},
\end{equation}
the components of the surface dynamics implied by Eqs.~(\ref{eq:ExpGenDyn}) and (\ref{eq:Xaxi}) are given by
\begin{align}
\partial_t r & = (v^u-q^u)r'-\frac{z'}{\sqrt{r'^2+z'^2}}v_n\label{eq:compdynr}\\
\partial_t z & = (v^u-q^u)z'+\frac{r'}{\sqrt{r'^2+z'^2}}v_n\label{eq:compdynz},
\end{align}
where primes $'$ denote partial derivatives with respect to~$u$ and azimuthal flows $v^{\phi}$ do by construction not contribute. Reparameterizing $\left\{r'(u,t),z'(u,t)\right\}\rightarrow\left\{(h(u,t),\psi(u,t)\right\}$, where $h=\sqrt{r'^2+z'^2}$ and $\psi$ is determined by
\begin{align}
r' &= h \cos\psi\label{eq:dur} \\
z' &= -h \sin\psi\label{eq:duz},
\end{align}
the shape dynamics Eqs.~(\ref{eq:compdynr}), (\ref{eq:compdynz}) become
\begin{align}
\partial_t r & = (v^u-q^u)h\cos\psi+v_n \sin\psi\label{eq:dtr}\\
\partial_t z & = (q^u-v^u)h\sin\psi+v_n\cos\psi\label{eq:dtz}.
\end{align}
Note that $\psi$ can be geometrically interpreted as the tangent angle~(Fig.~\ref{fig:SurfSketch})~\cite{seif91}, and is related to the meridional curvature tensor component by~$C^{\,u}_{u}=\psi'/h$. We find from the definition of $h$ and Eqs.~(\ref{eq:dur})--(\ref{eq:dtz}) that
\begin{align}
\partial_t h & = [(v^u-q^u)h]'+v_n\psi'.\label{eq:dth}
\end{align}
The significance of Eq.~(\ref{eq:dth}) for this work is as follows: From \hbox{$h=|\partial_u\mathbf{X}|$}, we see that $h(u,t)$ mediates via
\begin{equation}\label{eq:sfromh}
s(u,t)=\int_0^ud\bar{u}\,h(\bar{u},t)    
\end{equation}
a relationship between the fixed mesh coordinates~$u$ used to parameterize numerical solutions and the physical arc length $s(u,t)$ along the meridional outline~\cite{Mietke2019Self}. Consequently, the numerical resolution of the physical surface directly depends on the dynamic properties~of~$h(u,t)$. 

\subsection{Scaling Eulerian-Lagrangian parameterization}\label{sec:SLE}
For purely Eulerian ($q^u=v^u$) or Lagrangian ($q^u=0$) parameterizations the dynamics of~$h(u,t)$ Eq.~(\ref{eq:dth}), and consequently the resolution of the physical surface, is slave to the local surface flows~$\mathbf{v}$ (Tab.~\ref{tab:boundary}). To mitigate this lack of control, we choose a different approach in which the coordinate flow field~$q^u$ is determined by the constraint that the map $h(u,t)=h(t)$ used in Eq.~(\ref{eq:sfromh}) has to remain spatially constant and thus becomes fully independent of local surface flows or deformations. If this can be achieved, the resulting parameterization will in general be neither Eulerian nor Lagrangian and $L(t)=s(1,t)=h(t)$, where $L(t)$ is the total length of the meridional surface outline. Hence, $s(u,t)=L(t)u$ and $h(t)$ becomes a simple scaling factor that translates numerical mesh to physical arc length coordinates. We will refer to the resulting parameterization as scaling Lagrangian-Eulerian (SLE) parameterization.

To show a coordinate flow $q^u$ for such an SLE always exists, we impose $h'=0$ in Eq.~(\ref{eq:dth}) and solve for $q^u$, which yields
\begin{align}
q^u(u,t)&=v^u(u,t)+\frac{1}{h(t)}\int_0^ud\bar{u}\,v_n(\bar{u},t)\psi'(\bar{u},t)\nonumber\\
&+c_1(t)+c_2(t)u,  \label{eq:qusol}
\end{align}
where $c_1(t),c_2(t)$ are integration constants that depend on the boundary conditions for $q^u$ and $v^u$. The coordinate flow $q^u$ given in Eq.~(\ref{eq:qusol}) specifies the SLE parameterization and leads with Eq.~(\ref{eq:dth}) to
\begin{align}
\partial_th&=\left.\left(v^u-q^u\right)h\right|_0^1+\int_0^1du\,v_n\psi'.\label{eq:dtSEL2}
\end{align}
% Integration constants are determined from suitable boundary conditions for $q^u$ as
% \begin{align}
% c_1&=q^u(0)-v^u(0)\label{eq:c1}\\
% c_2&=-\left.\left(v^u-q^u\right)\right|_0^1-\frac{1}{h}\int_0^1du\,v_n\psi'.\label{eq:c2}
% \end{align}
% The coordinate flow Eq.~(\ref{eq:qusol}) is therefore Eulerian if the surface is static ($v_n=0$) and if both boundary points satisfy $q^u=v^u$. Before discussing boundary conditions, it is instructive to plug the SLE coordinate flow Eq.~(\ref{eq:qusol}) back into the dynamics of the coordinate map, Eq.~(\ref{eq:dth}), which yields
% \begin{align}
% \partial_th&=-[(c_1+c_2u)h]'\label{eq:dtSEL}\\
% &=\left.\left(v^u-q^u\right)h\right|_0^1+\int_0^1du\,v_n\psi'.\label{eq:dtSEL2}
% \end{align}
Equation~(\ref{eq:dtSEL2}) explicitly shows that an initially constant \hbox{$h=L_0$} will remain spatially constant over time and only changes as the length  $L(t)=h(t)$ of the meridional outline changes (see Fig.~\ref{fig:SurfSketch}) due to boundary flows (first term) and surface deformations (second term). Boundary conditions $q^u(u_b,t)$ for $u_b=0,1$ can in principle be chosen arbitrarily, but two natural choices are to treat a boundary point either as Lagrangian (\hbox{$q^u(u_b,t)=0$}) or Eulerian point ($q^u(u_b,t)=v^u(u_b,t)$). Both cases will be considered in the examples discussed below. 

\begin{table}
    \centering
    \begin{tabular}{cccc}
    \hline
        Parameterization & $q^u$ & $\partial_t h$ & References\\
        \hline
         Eulerian & $v^u$ & $ \psi' v_n $ & \cite{Mietke2019Self}\\
         Lagrangian & 0 & $ (v^u h)' + \psi'v_n$ &\cite{Turlier2014,Salbreux2023elife} \\
         SLE & s.t. $h' = 0$ & $dL/dt$ & This work\\
    \hline
    \end{tabular}
    \caption{Comparison of parameterizations classified by their coordinate flow $q^u$ [see Eq.~(\ref{eq:ExpGenDyn})]. $h(u,t)$ maps fixed mesh coordinates~$u$ to physical arc length coordinates~[see Eq.~(\ref{eq:dth})] and develops unpredictable flow- and shape-dependent inhomogeneities when purely Eulerian or Lagrangian parameterizations are used. The scaling Lagrangian-Eulerian (SLE) parameterization is independent of local flows $v^u$, deformations $v_n$ and curvature~$\psi'$ and amounts to a simple scaling between mesh and arc length coordinates.}
    \label{tab:boundary}
\end{table}

\section{Mechanics of deforming active fluid surfaces}\label{sec:ActSurf}
\subsection{Force and torque balance}\label{sec:FBTB}
The mechanical state of a curved surface is described by its tension and moment tensor, $\mathbf{T}=\mathbf{e}_i\otimes\mathbf{t}^i$ and \hbox{$\mathbf{M}=\mathbf{e}_i\otimes\mathbf{m}^i$}, where $\mathbf{t}^i=t^{ij}\mathbf{e}_j+t^i_n\mathbf{n}$ and \hbox{$\mathbf{m}^i=m^{ij}\mathbf{e}_j+m^i_n\mathbf{n}$} are tension and moment vectors whose components describe in-plane ($t^{ij},m^{ij}$) and out-of-plane or normal ($t^i_n,m^i_n$) contributions. In particular, for any line element of length $ds$ on the surface with an in-plane unit vector $\boldsymbol{\nu}=\nu_i\mathbf{e}^i$ orthogonal to it, vectors $\mathbf{F}=ds\,\boldsymbol{\nu}\cdot\mathbf{T}=ds\,\nu_i\mathbf{t}^i$ and $\boldsymbol{\tau}=ds\,\boldsymbol{\nu}\cdot\mathbf{M}=ds\,\nu_i\mathbf{m}^i$ yield the total force and the total torque, respectively, that act on the line element. Force and torque balance equations for the surface are given by~\cite{Guillaume2017}
\begin{align}
\div\left(\mathbf{T}\right)&=-\mathbf{f}^{\text{ext}}\label{eq:FB}\\
\div\left(\mathbf{M}\right)&=-\boldsymbol{\epsilon}:\mathbf{T}\label{eq:TB}.
\end{align}
Here, $\div\left(\mathbf{T}\right):=\mathbf{e}^i\cdot\partial_i\mathbf{T}$ denotes the in-plane divergence of the tension tensor (analog for $\mathbf{M}$), $\left(\boldsymbol{\epsilon}:\mathbf{T}\right)_{\alpha}=\epsilon_{\alpha\beta\gamma}T_{\beta\gamma}$, where $\epsilon_{\alpha\beta\gamma}$ is fully anti-symmetric Levi-Civita tensor (greek indices reference Cartesian components) and we do not include inertial forces or external torques. In the force balance Eq.~(\ref{eq:FB}), $\mathbf{f}^{\text{ext}}=f^{\text{ext}}_i\mathbf{e}^i+f^{\text{ext}}_n\mathbf{n}$ are external forces and we have neglected inertia, as we are interested in the typically overdamped dynamics of biological cells and tissues. In this work, we will consider external tangential friction forces
\begin{equation}\label{eq:fext}
f^{\text{ext}}_i=-\Gamma v_i
\end{equation}
with friction constant $\Gamma$. A general derivation of momentum conservation of curved surfaces described by an arbitrary Lagrangian-Eulerian parameterization is provided in the App.~\ref{app:momcons}. The torque balance Eq.~(\ref{eq:TB}) implies that inhomogeneous surface moments give rise to anti-symmetric components of the tension tensor~$\mathbf{T}$, which correspond to contributions from anti-symmetric in-plane tensions $t_{ij}$ and from normal forces~$t^i_n$. In the absence of moments, \hbox{$\mathbf{M}=0$}, $\mathbf{T}$ must be symmetric, which is equivalent to a symmetric in-plane tension tensor $t_{ij}$ and vanishing normal forces,~$t_n^i=0$. Center-of-mass velocities $\mathbf{v}$ that drive the surface dynamics Eq.~(\ref{eq:GenDyn}) will be determined by solving the force and torque balance equations~(\ref{eq:FB})~and~(\ref{eq:TB}). In the next step, we introduce the constitutive laws that relate tension $\mathbf{T}$ and moments $\mathbf{M}$ in the surface to its geometry and local center-of-mass velocity~$\mathbf{v}$.

\subsection{Constitutive laws}
We discuss in this work open and closed active fluid surface models that capture essential physical features of membranes, the cell cortex and epithelial tissues. In anticipation of formulating a dissipation functional for such surfaces, we introduce constitutive laws by closely following their systematic construction within the framework of irreversible thermodynamics~\cite{Guillaume2017,Julicher_2018}. We start by introducing the free energy density of the surface. The first contribution is given by the Helfrich free energy density~\cite{Helfrich1973}
\begin{equation}
\label{eq:helfrich}
    f_H=\gamma_H + f_{\kappa},
\end{equation}
where $\gamma_H$ denotes a passive isotropic surface tension and
\begin{equation}\label{eq:fkappa}
f_{\kappa}=\frac{1}{2}\kappa\left(2H-C_0\right)^2    
\end{equation}
includes a bending rigidity $\kappa$ , the spontaneous curvature $C_0$ and $H=\text{tr}(\mathbf{C})/2$ denotes the mean curvature. Second, we include the chemical free energy density $f_c(c_f,c_p)$ associated with surface concentrations $c_f$ and $c_p$ of chemical fuel and product species, respectively. In a biological system, these could be the reactant and product species involved in ATP hydrolysis~\cite{juli97,Prost2015naturephysics}. The chemical potential difference~$\Delta\mu=\mu_f-\mu_p$ with $\mu_n=\partial f_c/\partial c_n$ ($n=f,p$) provides the chemical energy to generate active stresses in the surface. The free energy of the whole surface~$\Omega$ is then given by
\begin{equation}\label{eq:Ffinal}
F=\int_{\Omega}dA(f_H+f_c).
\end{equation}
Such a free energy gives rise to well-defined, curvature-dependent equilibrium tensions and moments in the surface \cite{Guillaume2017,Mietke2019Minimal} that are in the following denoted by $\mathbf{T}_e$ and $\mathbf{M}_e$. 

To introduce dissipative contributions, we use the symmetrized strain rate tensor $S_{ij}=(v_{ij}+v_{ji})/2$, where $v_{ij}=\mathbf{e}_j\cdot(\partial_i \mathbf{v})$, and denote its traceless part by
\begin{equation}
\label{eq:defS}
\tilde{\mathbf{S}}=\mathbf{S}-\frac{1}{2}\text{tr}(\mathbf{S})\mathbf{G}
\end{equation}
with $\mathbf{S}=S_{ij}\mathbf{e}^i\otimes\mathbf{e}^j$. Note also that $\text{tr}(\mathbf{S})=\div(\mathbf{v})$. We consider a dissipative contribution to the tension tensor given~by
\begin{equation}
\label{eq:stressd}
    \mathbf{T}_d = 2\eta_s\tilde{\mathbf{S}} + \eta_b\,\div(\mathbf{v})\mathbf{G} + \xi \Delta \mu \mathbf{G},  
\end{equation}
where we have introduced shear viscosity and bulk viscosities $\eta_s$ and $\eta_b$, respectively, as well as an active isotropic tension $\xi\Delta \mu$ that captures the conversion of chemical free energy into mechanical work. We do not consider dissipative moments, such that the tension and moment tensors used in this work are given~by
\begin{align}
\mathbf{T}&=\mathbf{T}_e+\mathbf{T}_d\label{eq:Tgen}\\
\mathbf{M}&=\mathbf{M}_e.\label{eq:Mgen}
\end{align}
From Onsager's symmetry relations~\cite{onsa31,onsa31b,Julicher_2018}, we expect from Eq.~(\ref{eq:stressd}) an additional constitutive law for the thermodynamic variable conjugate to the chemical potential difference~$\Delta\mu$. Introducing such a constitutive law in way that is compatible with the formulation of a dissipation functional involves a crucial subtlety that will be discussed in the next step.

\section{Variational formulation}\label{sec:varform}
\subsection{Motivation}
For any given surface geometry and distribution of active tension $\sim\xi\Delta\mu$, force and torque balance Eqs.~(\ref{eq:FB}) and~(\ref{eq:TB}) with constitutive laws (\ref{eq:Tgen}) and (\ref{eq:Mgen}) determine in-plane flows and deformations $\mathbf{v}$. To characterize such an active surface theory systematically, we have to explore the space of non-trivial stationary geometries it gives rise to. To this end, we take inspiration from previous work on equilibrium membranes, for which phase diagrams of stationary equilibrium shapes have been determined using a variational formulation based on the free energy Eq.~(\ref{eq:helfrich}) with $f_c=0$~\cite{seifert1997configurations}. In these works, the free energy was complemented by Lagrange multipliers to impose the nonlinear geometric relations Eqs.~(\ref{eq:dur}) and (\ref{eq:duz}). While this yields a larger system of equations that is altogether still equivalent to the normal component of the force balance Eq.~(\ref{eq:FB}), it allows for direct calculations of stationary surface geometries described by energies like Eq.~(\ref{eq:helfrich}) via a much simpler boundary value problem. Here, we generalize this approach to out-of-equilibrium surfaces by constructing a Rayleigh dissipation functional $\mathcal{R}$ that imposes nonlinear geometric constraints, as well as an SLE parameterization, via suitable Lagrange multipliers and whose stationary points defined by the variation
\begin{equation}\label{eq:var}
\frac{\delta\mathcal{R}}{\delta\mathbf{v}}=0     
\end{equation}
are surfaces and flows that satisfy the force and torque balance Eqs.~(\ref{eq:FB}) and~(\ref{eq:TB}) for constitutive laws (\ref{eq:Tgen}) and (\ref{eq:Mgen}) %\todo{Are (31) and (32) considered as constitutive laws? I thought Eq. (30) is a constitutive law.}. 

\subsection{Free energy dynamics}
The dissipation functional consists of contributions from both the equilibrium properties of the surface, as well as its dissipative properties. The former result from changes of the free energy given in Eq.~(\ref{eq:Ffinal}), which follow from Eqs.~(\ref{eq:dtFfin2main}) and (\ref{eq:particle})~as
\begin{align}\label{eq:dtFequi}
\frac{dF}{dt}&=\int_{\Omega}dA\left(\frac{1}{\sqrt{g}}\partial_t(f_H\sqrt{g})-r_p\Delta\mu\right.\nonumber\\
&\hspace{1.3cm}+\gamma_e\div\left(\mathbf{v}\right)+J_n^{\text{ext}}\biggr)+\oint_{\partial\Omega}f_Hq^i\nu_ids,
\end{align}
where $r_p=-r_f$ denotes the rate at which the product species is produced due to fuel consumption, $\gamma_e=f_c-\mu_fc_f-\mu_pc_p$ is an equilibrium surface tension~\cite{Guillaume2017} and \hbox{$J_n^{\text{ext}}=J_{n,f}\mu_f+J_{n,p}\mu_p$} denotes the in-flux of chemical free energy due to reactions with the surrounding. We assume that the flux $J_n^{\text{ext}}$ is such that the chemical potential difference~$\Delta\mu$ on the surface is maintained constant in space and time. Consequently,~$\gamma_e$ only renormalizes the isotropic surface tension $\gamma_H$ and we denote the total passive tension by \hbox{$\gamma=\gamma_H+\gamma_e$}. The boundary term in Eq.~(\ref{eq:dtFequi}), which is a consequence of the ALE parameterization, contributes to boundary conditions on open surfaces when performing variation of~$\frac{dF}{dt}$. The Helfrich free energy dynamics in terms of an Eulerian parameterization has previously been derived in~\cite{Arroyo2009PRE}.

\subsection{Entropy production}
The dissipative tension tensor in Eq.~(\ref{eq:stressd}) can be obtained as an Onsager expansion of the entropy production $\Theta_{\text{int}}$ for an active isotropic fluid surface~\cite{Guillaume2017,GuillaumePRR2022}
\begin{equation}\label{eq:entrprod}
\Theta_{\text{int}}=\int dA\left(\mathbf{T}_d :{\bf S}+r_p\Delta\mu\right),  
\end{equation}
where each pair of conjugate thermodynamic variables contains one factor that is even ($\mathbf{T}_d$,$\Delta\mu$), and one that is odd ($\mathbf{S}$,$r_p$), under time-reversal. After designating for each pair one thermodynamic flux and one thermodynamic force, all fluxes can be expanded near thermodynamic equilibrium in terms of forces following Onsager theory~\cite{onsa31,onsa31b}. This procedure yields constitutive laws such as Eq.~(\ref{eq:stressd}), where $\eta_s,\eta_b$ and $\xi$ represent phenomenological Onsager coefficients~\cite{Guillaume2017,GuillaumePRR2022,Julicher_2018}. Given the most commonly made choice of treating $\Delta\mu$ as thermodynamic force~\cite{juli97}, as done in Eq.~(\ref{eq:stressd}), the constitutive law for the thermodynamic flux~$r_p$ reads
\begin{equation}\label{eq:constrp}
r_p=\Lambda\Delta\mu - \xi\,\div(\mathbf{v}).
\end{equation}
Onsager reciprocity relations require the cross-coupling with coefficient $-\xi$: Because flux $\mathbf{T}_d$ and force~$\Delta\mu$ have the same time reversal-symmetry -- the same is necessarily true for their cross-coupled conjugate partners $\div(\mathbf{v})$ and $r_p$ -- the coupling~$\xi$ is a reactive coefficient that must appear with opposite signs in the constitutive laws Eqs.~(\ref{eq:stressd}) and (\ref{eq:constrp})~\cite{Julicher_2018}. In the following, we will denote the entropy production Eq.~(\ref{eq:entrprod}) expressed in terms of constitutive laws Eqs.~(\ref{eq:stressd}) and (\ref{eq:constrp}) as functional $\Theta_{\text{int}}(\mathbf{v},\Delta\mu)$. Dissipative external forces, such as the effective friction introduced in Eq.~(\ref{eq:fext}), will additionally contribute to the entropy production with
\begin{equation}\label{eq:entrprodext}
\Theta_{\text{ext}}=-\int dA\mathbf{v}\cdot\mathbf{f}^{\text{ext}}.  
\end{equation}

\subsection{Flux-force choice and dissipation functional}\label{sec:Fluxchoice}
Unfortunately, variations of the \lq naive\rq\ entropy production Eq.~(\ref{eq:entrprod}) cannot recover the constitutive law Eq.~(\ref{eq:stressd}): Reactive couplings, such as the Onsager coefficient~$\xi$ in Eqs.~(\ref{eq:stressd}) and (\ref{eq:constrp}), do by construction not contribute to the entropy production. This can easily be verified by plugging Eqs.~(\ref{eq:stressd}) and (\ref{eq:constrp}) into the entropy production $\Theta_{\text{int}}(\mathbf{v},\Delta\mu)$ given in Eq.~(\ref{eq:entrprod}). We therefore adapt our strategy and try to formulate an equivalent set of constitutive laws exclusively in terms of dissipative Onsager coefficients. Here, this can be realized by choosing -- instead of~$\Delta\mu$ -- the chemical production rate $r_p$ as the thermodynamic force. We write the resulting constitutive laws as
\begin{align}
\label{eq:CR1}
\hat{\mathbf{T}}_d &= 2\eta_s\tilde{\mathbf{S}} + \left(\eta_b+\frac{\xi^2}{\Lambda}\right)\div(\mathbf{v})\mathbf{G} + \frac{\xi}{\Lambda}r_p\mathbf{G}\\
\Delta\hat{\mu} &= \frac{1}{\Lambda}r_p + \frac{\xi}{\Lambda}\div(\mathbf{v}),\label{eq:CR2}
\end{align}
where we have parameterized new Onsager coefficients that would appear such that Eqs.~(\ref{eq:CR1}) and (\ref{eq:CR2}) are equivalent to the desired constitutive laws Eq.~(\ref{eq:stressd}) and (\ref{eq:constrp}), i.e. $\hat{\mathbf{T}}_d=\mathbf{T}_d$ and $\Delta\hat{\mu}=\Delta\mu$. We denote the entropy production associated with constitutive laws Eqs.~(\ref{eq:CR1}) and (\ref{eq:CR2}) as functional
\begin{equation}\label{eq:ephat}
\hat{\Theta}_{\text{int}}(\mathbf{v},r_p)=\int dA\left(\hat{\mathbf{T}}_d :{\bf S}+r_p\Delta\hat{\mu}\right).
\end{equation}
Note that, consistent with Onsager reciprocity relations, the Onsager coefficient~$\xi/\Lambda$ now appears as a dissipative term with the same sign in both constitutive laws Eqs.~(\ref{eq:CR1}) and (\ref{eq:CR2}) and cross-couples flux-force pairs with opposite time-reversal signature. Consequently, it will contribute to the entropy production $\smash{\hat{\Theta}_{\text{int}}(\mathbf{v},r_p)}$. One can directly verify that if coefficients $\eta,\eta_b,\xi$ are such that the second law is respected in the original entropy production, i.e. $\Theta_{\text{int}}(\mathbf{v},\Delta\mu)\ge0$, then it will also be respected by $\smash{\hat{\Theta}_{\text{int}}(\mathbf{v},r_p)}$ in Eq.~(\ref{eq:ephat}).

%Furthermore, the constraint imposed by the second law of thermodynamics, $\smash{\hat{\Theta}_{\text{int}}(\mathbf{v},r_p)\ge0}$, is fully compatible with the conditions for the Onsager coefficients $\eta,\eta_b,\xi$ and $\Lambda$ we obtain from $\Theta_{\text{int}}(\mathbf{v},\Delta\mu)\ge0$ [see Eqs.~(\ref{eq:stressd}),(\ref{eq:entrprod}),(\ref{eq:constrp})]. Further, $\Theta_{\text{int}}(\mathbf{v},\Delta\mu)\ge0$ implies $\Lambda>0$ and $\eta_b\ge0$. Hence, $1/\Lambda>0$ and $\eta_b+\xi^2/\Lambda\ge0$, which is also independently required by the second law for $\hat{\Theta}_{\text{int}}(\mathbf{v},r_p)$ and Onsager coefficients parameterized as in Eqs.~(\ref{eq:CR1}) and (\ref{eq:CR2}). 

It has been pointed out that the designation of thermodynamic fluxes and forces is essentially arbitrary~\cite{meix43,katch65} and the physics described by the resulting constitutive laws will be equivalent in the thermodynamic limit for every choice. It is therefore not surprising that the transformation of the constitutive laws into Eqs.~(\ref{eq:CR1}) and (\ref{eq:CR2}), based on a modified flux-force pair choice, exists. The fact that associated entropy productions take different forms for different flux-force choices and only some choices yield a suitable dissipation functional does also not contradict any physical principle: In contrast to a principle such as energy minimization, the minimization of dissipation invoked by Rayleigh functionals is not a first physical principle and the existence of a suitable functional for any set of constitutive laws describing an out-of-equilibrium system is not guaranteed~\cite{katch65}. However, for constitutive laws that are linear in thermodynamic forces, Onsager relations guarantee an invertible coefficient matrix and thus ensure the existence of equivalent constitutive laws in terms of exclusively dissipative phenomenological coefficients. 

The total dissipation rate used in this work is given by
\begin{equation}\label{eq:disspart}
\mathcal{D}(\mathbf{v},r_p)=\frac{1}{2}\left(\hat{\Theta}_{\text{int}}+\Theta_{\text{ext}}\right),
\end{equation}
where $\hat{\Theta}_{\text{int}}(\mathbf{v},r_p)$ denotes the entropy production given in Eq.~(\ref{eq:ephat}) with $\hat{\mathbf{T}}_d$ and $\Delta\hat{\mu}$ given in Eqs.~(\ref{eq:CR1}) and~(\ref{eq:CR2}), respectively, as well as $\Theta_{\text{ext}}$ given in Eq.~(\ref{eq:entrprodext}). Together with contributions from the free energy in Eq.~(\ref{eq:dtFequi}), we arrive at the final Rayleigh dissipation functional
\begin{equation}
\label{eq:rayleigh}
\mathcal{R}(\mathbf{v},r_p)=\frac{dF}{dt}+\mathcal{D}.
\end{equation}
We show in~App.~\ref{app:FBequi} that variational equations (\ref{eq:var}) derived from $\mathcal{R}(\mathbf{v},r_p)$ correspond exactly to force and torque balance equations Eqs.~(\ref{eq:FB}), (\ref{eq:TB}) with tension and moments given by Eqs.~(\ref{eq:Tgen}), (\ref{eq:Mgen}). Variation $\delta\mathcal{R}/\delta r_p=0$ yield the remaining constitutive law Eq.~(\ref{eq:constrp}). 

\subsection{Enforcing geometric relations and SLE parameterization via Lagrange multipliers}\label{sec:enfsle}
The Rayleigh functional $\mathcal{R}(\mathbf{v},r_p)$, Eq.~(\ref{eq:rayleigh}) can now be supplemented by suitable Lagrange multipliers that dynamically impose nonlinear geometric relations and the SLE parameterization introduced above (Sec.~\ref{sec:SLE}). This will provide us with an approach to directly solve for stationary surface geometries and flows, and to compute the shape dynamic Eq.~(\ref{eq:GenDyn}) in an implicit fashion. To discuss this in detail, we express Eq.~(\ref{eq:rayleigh}) on an axisymmetric surface
\begin{equation}
\label{eq:rayleighaxi}
\mathcal{R}(\mathbf{v},r_p)=2\pi\int_0^1du\,R(\mathbf{v},r_p)+2\pi\left.(f_Hq^urh)\right|_0^1,
\end{equation}
where the explicit form of $R(\mathbf{v},r_p)$ is provided in App.~\ref{app:FBequi} [see Eq.~(\ref{eq:rayldens})]. Equation~(\ref{eq:rayleighaxi}) represents a functional $R[v^u,v^{\phi},v_n,\partial_t r,\partial_t h,\partial_t\psi,q^u,r_p]$. For Eulerian or Lagrangian parameterizations $\partial_t r$, $\partial_t h$ and $\partial_t\psi$ can be expressed in terms of flow velocities (App.~\ref{app:relpara}) and the coordinate flow~$q^u$ becomes fixed, such that the a functional is reduced to $R[v^u,v^{\phi},v_n,r_p]$. While variations with respect to velocity components do recover force and moment balance equations~(App.~\ref{app:vareul}), it disconnects from the shape dynamics in the embedding space and excludes explicit control over the mapping $h(u,t)$ between fixed mesh coordinates~$u$ and physical space [see Eq.~(\ref{eq:sfromh})]. However, imposing the SLE parameterization by directly fixing the coordinate flow~$q^u$ to Eq.~(\ref{eq:qusol}) leads to impractical integro-differential equations when performing the variation and is therefore also not feasible.
\begin{table}
    \centering
    \begin{tabular}{cc}
    \hline
        Variation & Implied ODE\\         
        \hline
         $\partial_t r$ & $\alpha' = \cdots $ \\
         $\partial_t z$ & $\beta' = \cdots $ \\
         $\partial_t\psi$ & $\psi'' = \cdots$  \\
         $\partial_t h$ &    $\zeta' = \cdots $ \\
         $\partial_t\alpha$ & $r' = \cdots $ \\
         $\partial_t\beta$ & $z' = \cdots $   \\
         $\partial_t \zeta$ & $h' = 0\hspace{0.32cm}$      \\
         $v^u$ & $\gamma'_H= \cdots $ / $(v^u)'' = \cdots $ \\
         $v^{\phi}$ & $(v^{\phi})'' = \cdots $\\
         $\gamma_H$ & $(v^u)' = \cdots $ / n.a.\\   
    \hline
    \end{tabular}
    \caption{List of fields with respect to which the dissipation functional \hbox{$\bar{\mathcal{R}}=\mathcal{R}+\mathcal{L}$}, with $\mathcal{R}$ and $\mathcal{L}$ given in Eqs.~(\ref{eq:rayleighaxi}) and (\ref{eq:Lmultiplier}), is varied and the differential equations implied by the corresponding variations. For an open incompressible surface variation w.r.t. the tension $\gamma_H(u)$ implies an additional ODE, that is not present in the compressible surface (n.a.). }
    \label{tab:equationsactive}
\end{table}
To sidestep an explicit specification of the coordinate flow~$q^u$ entirely, while still imposing an SLE parameterization, we proceed as follows. From the shape dynamics of an axisymmetric surface, Eqs.~(\ref{eq:dtr}) and (\ref{eq:dtz}), we find
\begin{align}
v_n&=\sin\psi\partial_tr+\cos\psi\partial_tz\\
q^u&=v^u+h^{-1}\left(\sin\psi\partial_tz-\cos\psi\partial_tr\right).\label{eq:quryv}
\end{align}
Using those identities to eliminate $v_n$ and $q^u$ from the Rayleigh functional leads to remaining dependencies $R[v^u,v^{\phi},\partial_t r,\partial_t z,\partial_t h,\partial_t\psi,r_p]$. Importantly, this substitution expresses two velocities, $v_n$ and $q^u$, in terms of three other velocities $\partial_tr,\partial_tz$ and~$v^u$. The relationship between $v_n$, $\partial_tr,\partial_tz$ and~$v^u$ becomes only unique once the coordinate flow~$q^u$ has been specified by a suitable constraint. This constraint is naturally given by $h'=0$, which we know leads to the SLE coordinate flow derived in Eq.~(\ref{eq:qusol}). However, the coordinate map~$h$ depends itself on $r,z$, $\psi$ via the nonlinear geometric identities Eqs.~(\ref{eq:dur}) and~(\ref{eq:duz}) and can therefore not be treated as an independent variable we would be allowed to constrain. To resolve this, we generalize an idea from work on passive membranes~\cite{seif91} and introduce Lagrange multipliers instead to impose the geometric identities Eqs.~(\ref{eq:dur}) and Eqs.~(\ref{eq:duz}). The full set of constraint terms is given~by
\begin{equation}
\label{eq:Lmultiplier}
    \mathcal{L} = 2\pi\int_0^1du\,\partial_t\left[\zeta h'+\alpha(r'-h \cos\psi)+\beta(z'+h\sin\psi)\right],
\end{equation}
where $\alpha$, $\beta$ and $\zeta$ represent Lagrange multipliers that dynamically enforce geometric relations Eqs.~(\ref{eq:dur}) and~(\ref{eq:duz}) and implicitly impose the SLE parameterization Eq.~(\ref{eq:qusol}). The final  constrained Rayleigh dissipation functional is then given by
\begin{equation}\label{eq:Rbar}
\bar{\mathcal{R}}=\mathcal{R}+\mathcal{L},   
\end{equation}
with $\mathcal{R}$ and $\mathcal{L}$ given in Eqs.~(\ref{eq:rayleighaxi}) and (\ref{eq:Lmultiplier}), respectively. This functional is varied with respect to the independent fields \hbox{$\Phi\in\left\{v^u, v^{\phi}, \partial_t r,\partial_t z, \partial_t h, \partial_t \psi, \partial_t\alpha, \partial_t\beta, \partial_t \zeta\right\}$} to determine equations of motion. The structure of the resulting system of ordinary differential equations (ODEs) obtained from
\begin{equation}\label{eq:Funcvar}
\frac{\delta\bar{\mathcal{R}}}{\delta\Phi}=0    
\end{equation}
is indicated in Tab.~\ref{tab:equationsactive}. Their explicit form is derived in App.~\ref{app:varsle}, where we also show that the resulting ODE system is again equivalent to the force balance equations of the active fluid surface. The bare ODE system yields for both incompressible and compressible fluid surfaces 12 first order equations. It must therefore be complemented by 12 boundary conditions, which are listed in Table~\ref{tab:boundaryactive} and can be derived from the boundary terms of the functional variation Eq.~(\ref{eq:Funcvar}) (App.~\ref{app:boundary}). In cases where closed surfaces with a conserved enclosed volume $V_0$ are considered (see Secs.~\ref{sec:SOsurf1}, \ref{sec:SOsurf2}) this ODE system is complemented by an equation $V'(u) = \pi r^2 h \sin\psi,$, which is by construction satisfied by the cumulative volume $V(u)$~\cite{seif91}. The boundary condition $V(1) = V_0$ imposes the volume constraint. A second boundary condition, $V(0) = 0$, fixes a degree of freedom that can directly be included as an external normal force $f_n^{\text{ext}}=p$ and corresponds to the pressure difference $p$ between the enclosed volume and the surrounding.

\begin{table}
    \centering
    \begin{tabular}{ccc}
    \hline
        \multirow{2}{*}{Variation}  & \multicolumn{2}{c}{Boundary conditions} \\
         & Open surface & Closed surface\\
        \hline
         $\partial_t r$ & $r(0) = 0,\,r(1) = R_b$\hspace{0.25cm}  & $r(0) = 0,\,r(1) = 0$\\
         $\partial_t z$ & $\beta(0) = 0,\,z(1) = 0$ \hspace{0.27cm} & $\beta(0) = 0,\,\beta(1) = 0$ \\
         $\partial_t\psi$ & $\psi(0) = 0$, $\psi(1) =0\ \ \ $ & $\psi(0) = 0$, $\psi(1) = \pi$ \\
         $\partial_t h$  & \multicolumn{2}{c}{$\zeta(0) = 0,\,\zeta(1) = 0$}\\
         $v^u$ & $v^u(0) = 0,\,\gamma_H(1) = \gamma_0\ \ \ \ $ & $v^u(0) = 0,\,v^u(1) = 0$\\
         $v^{\phi}$ & \multicolumn{2}{c}{$v^{\phi}(0) = 0,\,v^{\phi}(1) = 0$} \\ 
    \hline
    \end{tabular}
    \caption{Boundary conditions for open and closed surfaces, which follow from the boundary terms of $\delta\bar{\mathcal{R}}=0$~(see App.~\ref{app:boundary}).}
    \label{tab:boundaryactive}
\end{table}

\subsection{Implicit dynamics and direct computation of stationary states}\label{sec:statsurf}
Having obtained a formulation that simultaneously evolves surface coordinates in the embedding space ($r,z$) and in-plane geometric properties ($\psi,h$) enables us now to conveniently - and robustly - perform dynamic simulations of deforming fluid surfaces, as well as to directly compute stationary surface geometries and flows. 

\subsubsection{Implicit time-stepping method for the surface dynamics}
For the ODEs listed in Tab.~\ref{tab:equationsactive} the right-hand side contains the time derivatives $\partial_t r, \partial_t z, \partial_t \psi$, which enter all equations by construction linearly. We can therefore design a fully implicit time integration scheme to evolve the surface shape by replacing
\begin{equation}
\begin{split}
    \partial_t r & = \frac{r(u,t)-r(u,t-\Delta t)}{\Delta t}\\
    \partial_t z & = \frac{z(u,t)-z(u,t-\Delta t)}{\Delta t}\\
    \partial_t \psi & = \frac{\psi(u,t)-\psi(u,t-\Delta t)}{\Delta t},\\
\end{split}
\end{equation}
and using $r(u,t-\Delta t), z(u,t-\Delta t), \psi(u,t-\Delta t)$ from the previous time-step. Solutions $[r(u,t),z(u,t),\psi(u,t),v_u,v_{\phi}]$ of the ODE system derived from the variational formulation then directly correspond to surface geometries and flows at the next time point. Throughout this work, we use $\Delta t = 4\times10^{-4}\tau_{\eta}$ and $\Delta t = 10^{-3}\tau_{\eta}$ for simulations of open (see Tab.~\ref{tab:para_membrane}) and closed surfaces (see Tab.~\ref{tab:para_surface}), respectively. We provide a MATLAB implementation~\cite{git} to perform these iterative computations using standard boundary value solver~\cite{bvpsolv01}.

%%%%%%%%%%%%%%%%%%
%%%%%%%%%%%%%%%%%%
\begin{figure*}[t]
	\includegraphics[width = 2.05\columnwidth]{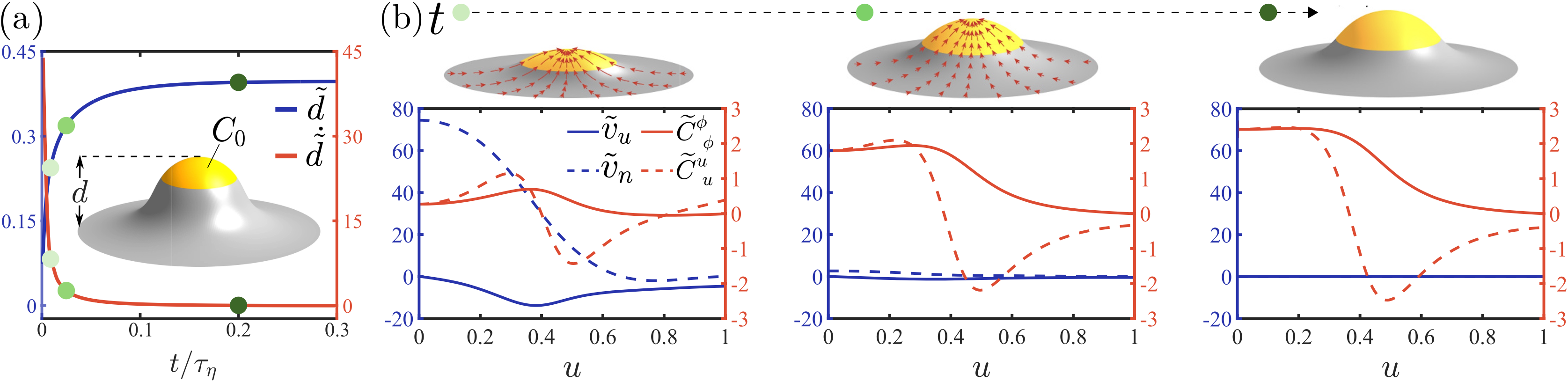}\vspace{-0.25cm}
	\caption{Dynamics of an open membrane with local spontaneous curvature coating [yellow patches, see Eq.~(\ref{eq:scpara})]. \textbf{(a)} Starting from a flat disk geometry, the membrane height $\tilde{d}=d/R_{b}$ (blue curve) grows in time $t$ due to the coating ($\tau_{\eta}=\eta_s R_b^2/\kappa$, $R_b$ is the fixed base radius). Red curve depicts growth speed. \textbf{(b)} Snapshots of evolving geometry with tangential components of local center-of-mass flow velocities indicated by red arrows (top row, Movie~1). Bottom row shows flow velocity and curvature tensor components: Tangential velocity $\smash{\tilde{v}_u=\bar{v}_u \tau_{\eta}/R_b}$, normal velocity $\smash{\tilde{v}_n=v_n \tau_{\eta}/R_b}$, azimuthal curvature $\smash{\tilde{C}_{\phi}^{\phi}=C_{\phi}^{\phi} R_{b}}$ and meridional curvature $\smash{\tilde{C}_u^u=C_u^u R_{b}}$. Last time~point shows final stationary geometry ($v_n=0$). Spontaneous curvature $\smash{\tilde{C}_0=\hat{C}_0R_b}$ is set to $\smash{\tilde{C}_0 = 9}$. All other parameters are listed in Table~\ref{tab:para_membrane}.}
	\label{fig:MembrDyn}
\end{figure*}
%%%%%%%%%%%%%%%%%%
%%%%%%%%%%%%%%%%%%

\subsubsection{Direct computation of stationary geometries and flows}\label{sec:dcomp}
While dynamic simulations are useful to explore the parameter-dependence of a given model, they are not practical to systematically characterize the space of non-trivial stationary surface geometries and flows or the bifurcation structure of this space. With the formulation derived above, we can directly compute these features by setting~instead
\begin{equation}
\begin{split}
    \partial_t r & = 0\\
    \partial_t z & = v_0\\
    \partial_t \psi & = 0,\\
\end{split}
\end{equation}
where $v_0$ is an additional parameter that represents a possible translational velocity of the surface along the symmetry axis. For open surfaces that are fixed at the base by boundary condition $z(1)=0$, we have $v_0=0$ (see Tab.~\ref{tab:boundaryactive}). For closed surfaces with asymmetric surface flows and in the presence of friction, $v_0$ is an unknown parameter determined as part of the boundary value problem. In the latter case, the boundary condition $z(0) + z(1) = 0$ is added to fix a reference frame that is co-moving with the surface~\cite{Mietke2019Self}. We test the stability of stationary solution by applying a small perturbation and using the resulting geometry as initial condition in dynamic simulations (SI~Fig.~\ref{fig:stabtest}). We note that the direct computation of stationary solutions, even those that may sill exhibit tangential flows, is only made possible by using the SLE parameterization. 

\section{Surface dynamics and shape spaces}
\subsection{Formation of membrane protrusion via internal and external forces}\label{sec:membr}

\begin{table}
    \centering
    \begin{tabular}{cccc}
    \hline
        \multicolumn{2}{c}{Parameter} & Value & Unit\\
        \hline 
        $\kappa$ & Bending rigidity &  1 & $\kappa$ \\  
        $\eta_s$ & Shear viscosity & 1  & $\eta_s$   \\
        $R_b$ & Base radius & 1 & $R_b$ \\
        $\tau_\eta$ & Time scale & $\eta_sR_b^2/\kappa$ & $\tau_\eta$\\
        $\Gamma$ & Friction & 0.25 & $\eta_s/R_b^2$ \\
        $\gamma_0$ & Boundary tension &  12.5 & $\kappa R_b^{-2}$\\
        $A_p$ & Area of coating & 2/25 & $2\pi R_b^2$\\
        $\sigma_p$ & Coating profile sharpness & 25 & $1/(2\pi R_b^2)$ \\
        $A_f$ & Area of external force & 2/250 & $2\pi R_b^2$ \\
        $\sigma_f$ & Force profile sharpness & 250 & $1/(2\pi R_b^2)$\\
    \hline        
    \end{tabular}
    \caption{Parameters used in open membrane examples shown in Figs.~\ref{fig:MembrDyn}~and~\ref{fig:MembrShapeSpace}, the characteristic time is defined as $\tau_{\eta}=\eta_s R_b^2/\kappa$.}
    \label{tab:para_membrane}
\end{table}

To illustrate the flexibility of our method and its capacity to characterize emergent shape spaces, we first study deformations of an open circular patch of membrane in response to internal and external forces, and neglect contributions from the chemical free energy [consequently $\xi\Delta\mu=0$ in Eq.~(\ref{eq:stressd})]. In all cases, we consider an incompressible membrane for which the isotropic tension $\gamma_H$ in the Helfrich free energy Eq.~(\ref{eq:helfrich}) becomes a Lagrange multiplier that has to be dynamically solved for (see Tab.~\ref{tab:equationsactive}). The free energy dynamics Eq.~(\ref{eq:dtFequi}) simplifies on the axisymmetric semi-open surface (Fig.~\ref{fig:SurfSketch}a) to 
\begin{equation}\label{eq:dFdtmembr}
\frac{dF}{dt}=2\pi\hspace{-0.15cm}\int_0^1\hspace{-0.15cm}du\left[\partial_t(rhf_{\kappa})+rh\gamma_{H}\div(\mathbf{v})\right]+\left.2\pi f_{\kappa}q^uhr\right|_{u=1},
\end{equation}
where $\div(\mathbf{v})$ and the Helfrich density $f_{\kappa}$ on axisymmetric surfaces are given in Eqs.~(\ref{eq:divvexp}) and (\ref{eq:HelfAxi}), respectively. 

\subsubsection{Local spontaneous curvature}
%%%%%%%%%%%%%%%%%%
%%%%%%%%%%%%%%%%%%
\begin{figure*}[t]
	\includegraphics[width = 2.05\columnwidth]{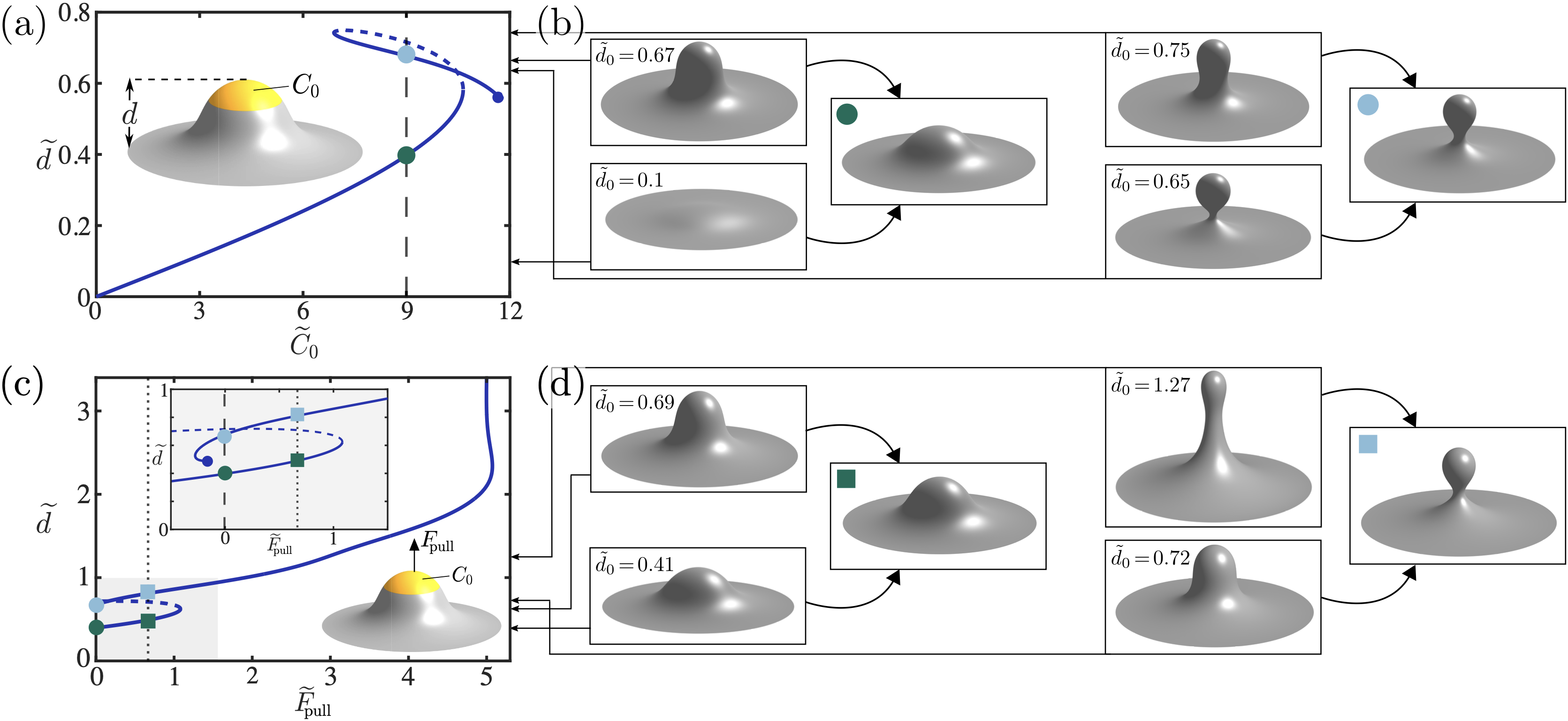}\vspace{-0.3cm}
	\caption{Nonlinear shape space analysis of open membranes. \textbf{(a)}~Height $\tilde{d}=d/R_b$ of stationary surfaces as function of spontaneous curvature $\smash{\tilde{C}_0 = \hat{C}_0R_b}$ in a finite-sized patch (yellow). Solutions obtained by direct computation of stationary geometries (Sec.~\ref{sec:dcomp}) reveal a first order transition due to a Gibbs loop with stable (solid blue line) and unstable (dashed blue line) branch regions. Solution branch terminates in limiting shape associated with a budding transition (dark blue circle). \textbf{(b)}~Depending on initial height $\tilde{d}_0$ and geometry, membranes relax for fixed spontaneous curvature ($\smash{\tilde{C}_0 = 9}$) to stationary shapes without (green circle) and with neck (blue circle). \textbf{(c)}~Stationary shape trajectory with $\smash{\tilde{C}_0 = 9}$ under additional application of external pulling forces $\smash{\tilde{F}_{\text{pull}}=F_{\text{pull}}R_b/\kappa}$ [see Eq.~(\ref{eq:pullforce})]. Inset shows zoom-in of gray region extended to \smash{$\tilde{F}_{\text{pull}}<0$}. The budding transition for this spontaneous curvature occurs when the membrane is \textit{pushed} (\smash{$\tilde{F}_{\text{pull}}<0$}). Branch regions without and with necks become fully disconnected in the ($\tilde{F}_{\text{pull}},\tilde{d}$)-space (inset), leading to a distinct first order transition near \smash{$\tilde{F}_{\text{pull}}\approx1$}. Pulling forces saturate following an additional first order transition near \smash{$\smash{\tilde{F}_{\text{pull}}\approx5}$}, consistent with ref.~\cite{julicher2002}. \textbf{(d)}~Representative examples of shape relaxations at fixed pulling force ($\tilde{F}_{\text{pull}}=0.7$) for different initial geometries (Movie~2). All other parameters are listed in~Table~\ref{tab:para_membrane}.}
	\label{fig:MembrShapeSpace}
\end{figure*}
%%%%%%%%%%%%%%%%%%
%%%%%%%%%%%%%%%%%%

We first assign a finite spontaneous curvature ``coating" to a sub-region of the membrane~(Fig.~\ref{fig:MembrDyn}a, inset), as can be induced by, for example, protein-patches on lipid membranes that play a role in clathrin-mediated endocytosis~\cite{hassinger2017design,Ma2021,walani2015endocytic}. We assume this coating occupies the same material points at all the times. Starting from an initial profile~\cite{agra09}
\begin{equation}\label{eq:scpara}
C_0(u) = \frac{\hat{C}_0}{2}\left(1-\tanh\left\{\sigma_p\left[
    A(u)-A_p\right]\right\}\right),
\end{equation}
where $A(u)=2\pi\int_0^udu'rh$ denotes the cumulative surface area up to mesh coordinate~$u$, the coating must therefore satisfy $\frac{dC_0}{dt} = 0$ or, explicitly,
\begin{equation}
\label{eq:dC0dt}
    \partial_t C_0 = -q^u\partial_uC_0,
\end{equation}
where $q^u$ is the SLE coordinate flow given in Eq.~(\ref{eq:quryv}). 

Starting from a flat geometry, surfaces with such a coating gradually bend and form protruding geometries~(Fig.~\ref{fig:MembrDyn}) as a result of the spontaneous curvature coating. At first, the normal velocity $v_n$ has its maxima at the center $u = 0$ and decays towards the base $u = 1$. Gradients of spontaneous curvature generate inhomogeneous in-plane tension (App.~\ref{app:equitens}), which leads to in-plane flows~$v_u$ that accompany the protrusion process (red arrows in Fig.~\ref{fig:MembrDyn}b). Growth eventually saturates, leading to a non-trivial stationary membrane geometry and vanishing center-of-mass flows (Fig.~\ref{fig:MembrDyn}b, right). 

We now characterize the space of emergent stationary geometries as a function of the local spontaneous curvature \hbox{$\tilde{C}_0=\hat{C_0}R_b$} systematically. The corresponding trajectories of stationary geometries are obtained by direct computation as described in Sec.~\ref{sec:dcomp}. This analysis reveals a non-monotonic, non-bijective behavior, known as Gibbs loop, of the stationary membrane height $d$ as a function of~$\tilde{C}_0$~(Fig.~\ref{fig:MembrShapeSpace}a). Consequently, for an intermediate range of local spontaneous curvature, three distinct stationary geometries are associated with a given spontaneous curvature imposed by the coating. Subsequent stability testing identifies two stable and one unstable region of the solution branch (solid and dashed lines, respectively, in Fig.~\ref{fig:MembrShapeSpace}a) in the degenerate parameter region. Such a degeneracy of membrane shapes was previously discussed in the context of a snap-through instability~\cite{hassinger2017design}. The trajectory of stationary surfaces terminates in a limiting geometry (dark blue circle in Fig.~\ref{fig:MembrShapeSpace}a) associated with a budding transition. Fixing $\tilde{C}_0 = 9$ (black vertical dashed line in Fig.~\ref{fig:MembrShapeSpace}a) and considering different geometries as initial conditions, the convergence to stationary surface geometries on the stable branch is illustrated in Fig.~\ref{fig:MembrShapeSpace}b. Stationary solutions indicate the separated stable regions correspond to protrusions with (without) a neck, defined as surfaces with (without) a region of negative Gaussian curvature (green and blue circles in Figs.~\ref{fig:MembrShapeSpace}a,b). The first order transition indicated by the Gibbs loop in the ($\tilde{C}_0$,$\tilde{d}$)-space is therefore a transition between these two basic protrusion geometries, consistent with previous observations~\cite{hassinger2017design,Ma2021}. An additional signature of this discontinuous transition is an effective shape hysteresis (see Fig.~\ref{fig:MembrShapeSpace}b): A further protruded surface ($\tilde{d}_0=0.67$) without neck may still relax to a flatter final state (green dot), while a slightly less protruded initial surface ($\tilde{d}_0=0.65$) can relax to a further extended geometry (blue dot) if an initial neck is present.

%%%%%%%%%%%%%%%%%%
%%%%%%%%%%%%%%%%%%
\begin{figure*}[t]
	\includegraphics[width = 2.05\columnwidth]{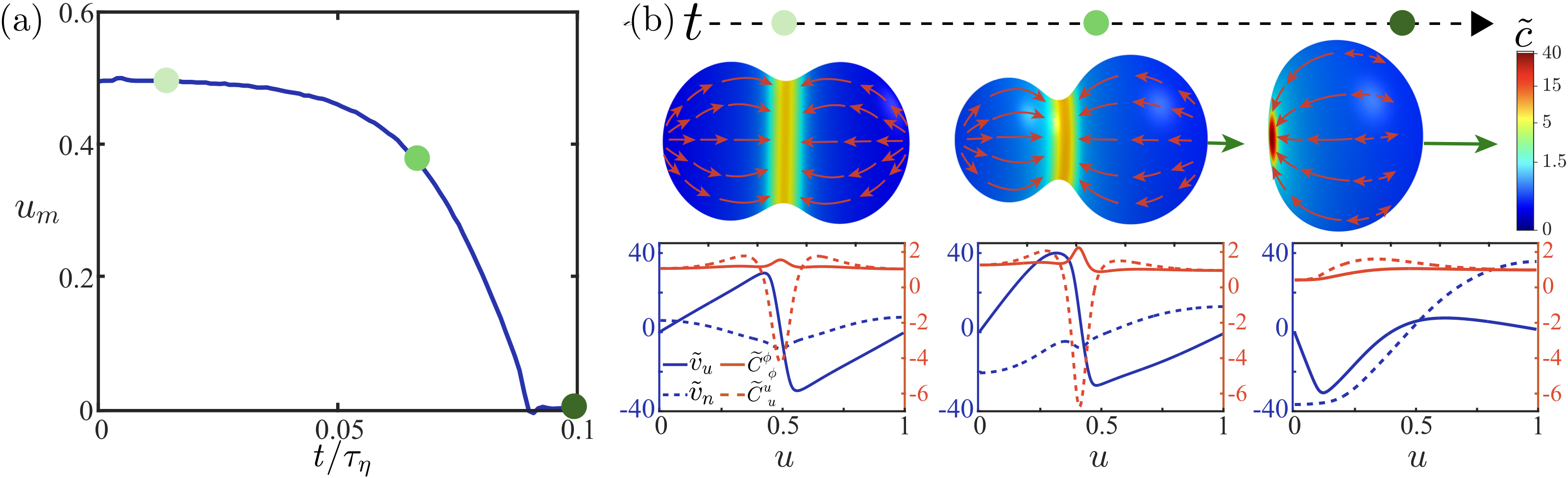}\vspace{-0.3cm}
	\caption{Self-organization dynamics, contractile ring slipping and propagation of a closed active isotropic fluid surface. \textbf{(a)}~Mesh coordinate~$u_m$ of maximal stress-regulator concentration over time ($\tau_{\eta}=\eta_b R_0^2/\kappa$) when initiating a centered contractile ring profile $c(u)_{t=0} = c_0[0.2 + 2\exp(-(u-1/2)^2/w^2)]$. The contractile ring remains initially centered but eventually slips to the pole ($u=0$). \textbf{(b)}~Snapshots of surface shapes with surface flow velocities indicated by red arrows, colour code indicates stress regulator concentration $\tilde{c}=c/c_0$. Last snapshot shows final surface geometry. Velocity ($\smash{\tilde{v}_u=\bar{v}_u \tau_{\eta}/R_0}$, $\smash{\tilde{v}_n=v_n \tau_{\eta}/R_0}$) and curvature ($\smash{\tilde{C}_{\phi}^{\phi}=C_{\phi}^{\phi} R_{0}}$, $\smash{\tilde{C}_u^u=C_u^u R_{0}}$) component profiles at corresponding time points are shown in the bottom row. Polar symmetry breaking leads to directed surface flows as previously found~\cite{Mietke2019Self,Mietke2019Minimal}. Here, due to friction with the surrounding [see Eq.~(\ref{eq:fext})], these flows lead to a translation of the final stationary shape. Green arrows indicate relative swimmer speed. Surface flow velocity arrows and velocity components are shown in the lab frame. As a result, $v_n\ne0$ in the last snapshot, even though the surface geometry is stationary. Parameters are listed in Table~\ref{tab:para_surface}.} 
	\label{fig:RingSlip}
\end{figure*}
%%%%%%%%%%%%%%%%%%
%%%%%%%%%%%%%%%%%%

\subsubsection{Pulling forces}
We next study membrane deformations under an external force density $\mathbf{f}^{\rm{ext}}= f^{\rm{ext}}_z\mathbf{e}_z$, while maintaining the spontaneous curvature coating. The external force density amounts to a total applied pulling force $F_{\text{pull}} = \int dAf^{\rm{ext}}_z$. The pulling force density is parameterized by 
\begin{equation}\label{eq:pullforce}
    f^{\rm{ext}}_z(u) = \frac{f_0}{2}\left(1-\tanh\left\{\sigma_f\left[
    A(u)-A_f\right]\right\}\right).
\end{equation}
In Eq.~(\ref{eq:pullforce}), $f_0$ is the magnitude of the applied force density, $A_f$ denotes the approximate total area of the region to which it is applied, and $\sigma_f$ is the inverse width of the transition between forced and unforced membrane regions. The impact of turgor pressure on such a configuration has previously been described~\cite{Ma2021}. Here, we are interested in the interplay of pulling forces and spontaneous curvature coating. Pulling forces in the absence of a spontaneous curvature coating give rise to a first-order transition after which membrane protrusions become tubular and keep growing at constant force~\cite{julicher2002}. This phenomenology is recovered in Fig.~\ref{fig:MembrShapeSpace}c at $\smash{\tilde{F}_{\text{pull}}\approx5}$, in quantitative agreement with results for singular point pulling forces obtained in ref.~\cite{julicher2002}, where a characteristic force \smash{$\hat{f}_0 = 2\pi \sqrt{\gamma_0\kappa}$} was used, while the characteristic force in our case is $\kappa/R_b\approx\hat{f}_0/5$ for the parameters used (see Tab.~\ref{tab:para_membrane}). However, we now find an additional first-order transition at smaller pulling forces ($\tilde{F}_{\text{pull}}\approx1$) when the spontaneous curvature coating is present. Closer inspection (inset in Fig.~\ref{fig:MembrShapeSpace}c) reveals that solution branches without and with necks are now fully disconnected in the \smash{($\tilde{F}_{\text{pull}},\tilde{d}$)}-space, while they form loops in the \smash{($\tilde{C}_0,\tilde{d}$)}-space when $\tilde{F}_{\text{pull}}=0$ (see Fig.~\ref{fig:MembrShapeSpace}a) signifying another spontaneous curvature-associated, but physically distinct, first order transition. We further find that the budding transition can occur for compressive forces ($\tilde{F}_{\text{pull}}<0$, see inset Fig.~\ref{fig:MembrShapeSpace}c, dark blue circle) when a spontaneous curvature coating is present. Exemplary transitions under fixed pulling force ($\tilde{F}_{\text{pull}}=0.7$, black vertical dashed line in Fig.~\ref{fig:MembrShapeSpace}c) between different initial conditions and corresponding stationary states are shown in Fig.~\ref{fig:MembrShapeSpace}d, which confirms the disconnected solution branches in the \smash{($\tilde{C}_0,\tilde{d}$)}-space indeed correspond to stationary surfaces without and with neck (green and blue squares in Figs.~\ref{fig:MembrShapeSpace}c,d), as suggested by their association with the corresponding branches of the Gibbs loop when $\tilde{F}_{\text{pull}}=0$ (circles in Fig.~\ref{fig:MembrShapeSpace}a,c).

\subsection{Spontaneous polarization, propagation and guided division of closed active fluid surfaces}\label{sec:SelfOrgaSurf}
We now include chemically regulated active stresses $\sim\xi\Delta \mu$ as introduced in the tension tensor Eq.~(\ref{eq:stressd}) and investigate self-organized deformations and emergent shape spaces of closed active fluid surface. Following previous work~\cite{Bois2011PRL,Mietke2019Self,Mietke2019Minimal,bona22}, we consider a concentration field $c$ that represents force-generating molecules and satisfies the continuity equation Eq.~(\ref{eq:particle}). Specifically, we use a diffusive flux $j_i = - D \partial_ic$ with diffusion constant $D$ and a normal flux $J_{n} = -k(c - \hat{c}_0)$ describing the exchange of molecules with the surrounding at a rate $k$ to maintain a reference concentration $\hat{c}_0$ on the surface. The dynamics of stress regulators reads
\begin{equation}\label{eq:cdt}
    \frac{dc}{dt}+\div(\mathbf{v})c=D\Delta c-k(c-\hat{c}_{0}),
\end{equation}
where $\frac{dc}{dt}=\partial_tc+q^i\partial_ic$, $\Delta$ is the Laplace-Beltrami operator on the curved surface and an expression for $\div(\mathbf{v})$ is provided in Eq.~(\ref{eq:divvexp}). We will explore a general scenario in which the reference concentration $\hat{c}_0$ profile can correspond to a centered contractile ring, i.e. 
\begin{equation}
\label{eq:c0dyn}
    \hat{c}_{0}(u) = c_0\left[1 + \Delta_c\exp\left(-\frac{\left(u-\frac{1}{2}\right)^2}{w^2}\right)\right].
\end{equation}
The cases $\Delta_c=0$ and $\Delta_c>0$ will be discussed in Secs.~\ref{sec:SOsurf1} and \ref{sec:SOsurf2}, respectively. In the context of a cell cortex, an externally imposed profile of the form Eq.~(\ref{eq:c0dyn}) can result from biochemical signaling originating from the spindle~\cite{ther07,neli15}. We assume the active isotropic stress depends on the regulator concentration according to~\cite{Mietke2019Self}
\begin{equation}\label{eq:stresssc}
\xi\Delta\mu(c)=(\xi\Delta\mu)_0\frac{c^2}{c^2+c_s^2},
\end{equation}
where $c_s$ denotes the saturation density.

%%%%%%%%%%%%%%%%%%
%%%%%%%%%%%%%%%%%%
\begin{figure*}[t]
	\includegraphics[width = 2.05\columnwidth]{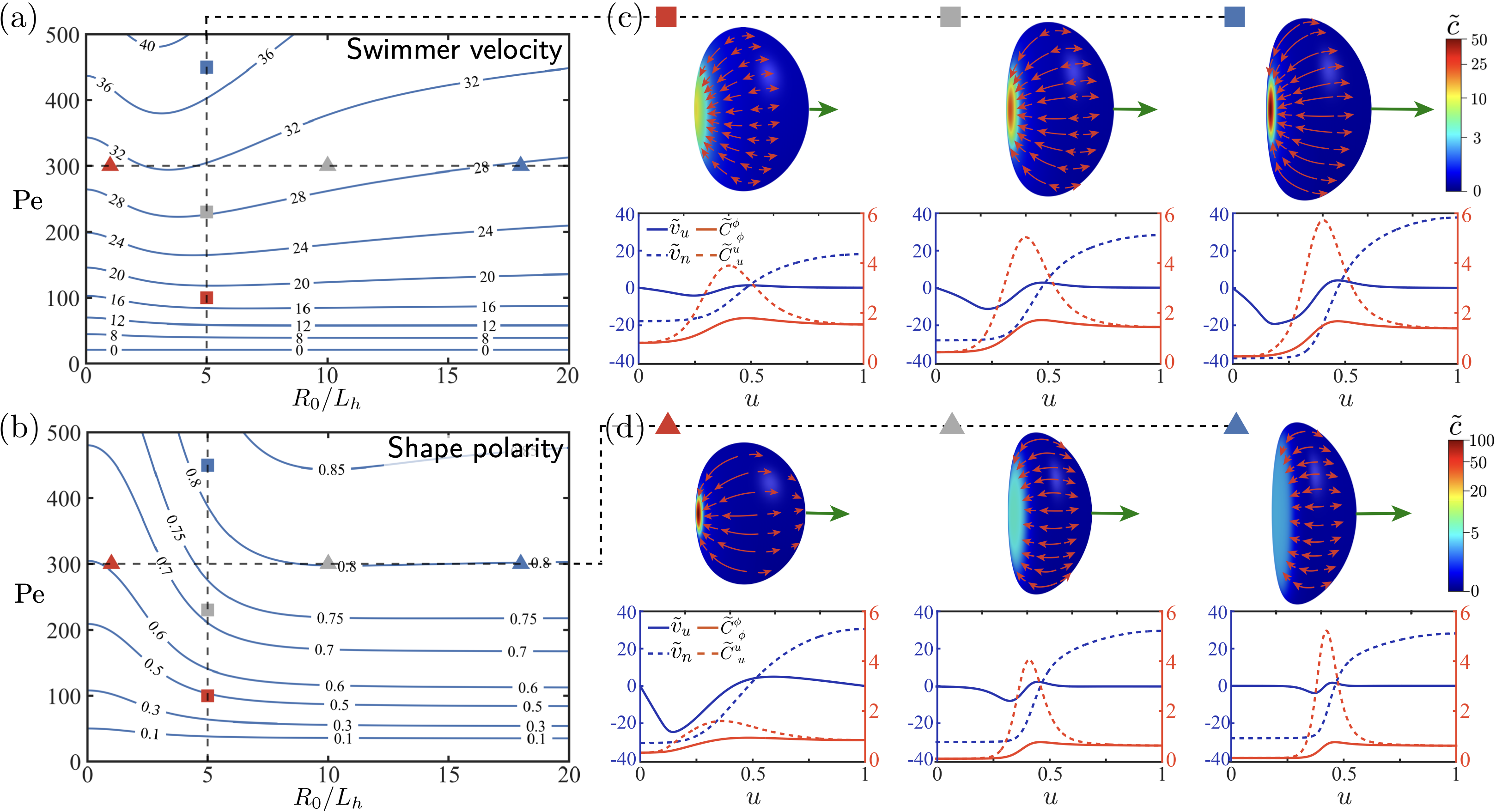}\vspace{-0.2cm}
	\caption{Shape space characterization of polarized stationary surfaces. \textbf{(a)}~Contour plot of swimmer velocity as function of P\'eclet number $\text{Pe}=(\xi\Delta\mu)_0R_0^2/(D\eta_b)$ and inverse relative hydrodynamic length $R_0/L_h$ ($\smash{L_h=\sqrt{\eta_b/\Gamma}}$). Contours were obtained by direct computation of nonlinear stationary surface solutions using the approach described in Sec.~\ref{sec:statsurf}. \textbf{(b)}~Surface shape polarity $P = 1 - H(0)/H(1)$ (contours), where $H(u)=C_k^{\,k}/2$ is the mean curvature in same parameter space as (a). \textbf{(c)}~Exemplary stationary solutions when increasing P\'eclet number at fixed hydrodynamic length ($R_0/L_h=5$): Swimming speed increases as a result of increased tangential flow velocities (Movie~3). \textbf{(d)}~At fixed P\'eclet number ($\text{Pe}=300$), a decreasing hydrodynamic length increases the geometric shape polarity until it saturates. Panel shows exemplary stationary solutions with tangential flow velocities (red arrows), surface propagation speed (green arrows) and concentration patterns (color-code). We used $k\tau_{\eta}=9$ and $c_s/c_0=1$, all other parameters are listed in Table~\ref{tab:para_surface}.}
	\label{fig:swimana}
\end{figure*}
%%%%%%%%%%%%%%%%%%
%%%%%%%%%%%%%%%%%%

\subsubsection{Spontaneous polarization and propagation}\label{sec:SOsurf1}
\begin{table}
    \centering
    \begin{tabular}{cccc}
    \hline
        \multicolumn{2}{c}{Parameter} & Value & Unit\\
        \hline 
        $\kappa$ & Bending rigidity &  1 & $\kappa$ \\         
        $\eta_b$ & Bulk viscosity & 1  & $\eta_b$   \\
        $\eta_s$ & Shear viscosity & 1 & $\eta_b$  \\      
        $R_0$ & Spherical radius & 1 & $R_0$ \\
        $\tau_\eta$ & Time scale & $\eta_bR_0^2/\kappa$ & $\tau_\eta$\\
        $(\xi\Delta \mu)_0$ & Active contractility & 450 &$D\eta_b/R_0^2$\\
        $\Gamma$ & Friction & 0.09 & $\eta_b/R_0^2$ \\
        $\gamma=\gamma_H+\gamma_e$ & Passive tension & 9 &$D\eta_b/R_0^2$\\
        $D$ & Diffusion constant & 1 & $R_0^{2}/\tau_\eta$\\
        $k$ & Turnover rate & 45 & $\tau_\eta^{-1}$\\
        $c_0$ & Steady state concentration & 1 & $c_0$ \\
        $c_s$ & Saturation concentration & 10 & $c_0$ \\
        \multirow{ 2}{*}{$w$} & Preferred concentration& \multirow{ 2}{*}{0.02} & \multirow{ 2}{*}{1} \\
         & profile width & \\
    \hline
    \end{tabular}
    \caption{Parameters used in closed active fluid surface examples shown in Figs.~\ref{fig:RingSlip}--\ref{fig:furrowshsp} (parameters $k$ and $c_s$ are only used in Figs.~\ref{fig:furrowdyn},~\ref{fig:furrowshsp}). The characteristic time in these simulations is defined as $\tau_{\eta}=\eta_b R_0^2/\kappa$. Characteristic length scale $R_0$ is related to the enclosed conserved volume by $V = 4\pi R_0^3/3$.}
    \label{tab:para_surface}
\end{table}

%%%%%%%%%%%%%%%%%%
%%%%%%%%%%%%%%%%%%
\begin{figure*}[t]
	\includegraphics[width = 2.05\columnwidth]{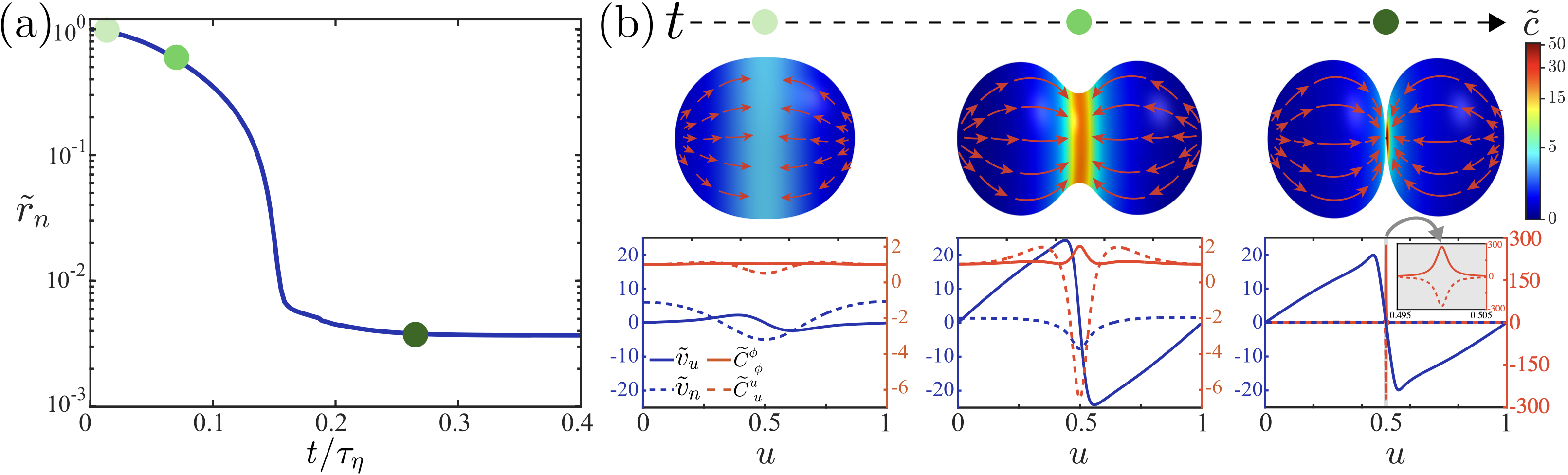}\vspace{-0.2cm}
	\caption{Dynamics of guided furrow formation and emergence of extreme localized curvature (Movie~4). \textbf{(a)} Evolution of the neck radius $\tilde{r}_n=r_n/R_0$ with $r_n:=r|_{u=1/2}$ from an initially spherical surface ($\tilde{r}_n=1$) and homogeneous concentration ($c(u)_{t=0}=c_0$), where force generator dynamics Eq.~(\ref{eq:cdt}) includes an inhomogeneous pattern of preferred concentration $\hat{c}_0(u)$~[Eq.~(\ref{eq:c0dyn})]. The magnitude of the inhomogeneity is set to $\Delta_c = 4$. $\tau_{\eta}=\eta_bR_0^2/\kappa$. A finite neck radius remains due to bending rigidity~$\kappa$. \textbf{(b)} Snapshots of surface shapes with surface flow velocities indicated by red arrows, colour code indicates stress-regulator concentration $\tilde{c}=c/c_0$. Bottom panels show velocity ($\smash{\tilde{v}_u=\bar{v}_u \tau_{\eta}/R_0}$, $\smash{\tilde{v}_n=v_n \tau_{\eta}/R_0}$) and curvature tensor ($\smash{\tilde{C}_{\phi}^{\phi}=C_{\phi}^{\phi} R_{0}}$, $\smash{\tilde{C}_u^u=C_u^u R_{0}}$) component profiles. An extreme localized curvature ($\Delta\tilde{C}_u^{\,u}/\Delta u\approx 5\times10^4$) arises in the neck region $u=1/2$ (last time point, inset), which our approach can robustly simulate using a standard boundary value solver. All other parameters are listed in Table~\ref{tab:para_surface}.}
	\label{fig:furrowdyn}
\end{figure*}
%%%%%%%%%%%%%%%%%%
%%%%%%%%%%%%%%%%%%

We first discuss the case $\Delta_c=0$ in Eq.~(\ref{eq:c0dyn}) for which the reference concentration $\hat{c}_0=c_0$ in Eq.~(\ref{eq:cdt}) is constant. Using this model, it has been shown in the absence of friction that a spherical surface with uniform distribution of stress-regulator undergoes a mechano-chemical shape instability above a critical active stress~\cite{Mietke2019Self}. As a result of this instability, chemical patterns and surface geometry break symmetry along the $z$-axis, eventually reaching a steady state that is polarized. This polarization dynamics still occurs in the presence of friction, and even if alternative patterns are imposed initially, as we demonstrate in~Fig.~\ref{fig:RingSlip}. There, we use a centered ($u = 1/2$) ring-like concentration pattern as initial condition and follow the position of the concentration maximum over time~(Fig.~\ref{fig:RingSlip}a). While such ring patterns remain initially centered, they eventually move towards one of the surface's poles ($u=0,1$, see Fig.~\ref{fig:SurfSketch}). Surface snapshots of this process and quantitative information on curvature and flow speeds are provided in Fig.~\ref{fig:RingSlip}b. The ring does initially lead to an ingression before it slips, consistent with geometric minimal models of cell division~\cite{sedz11,dead23}, and a polar concentration pattern and surface geometry remain. Due to the friction, and in the presence of a polarized surface state, surfaces will propagate in the lab frame. Green arrows in Fig.~\ref{fig:RingSlip} indicate the relative magnitude of propagation speeds as ring-slipping occurs, which shows a clear correlation between polarization and propagation speed. The physics of this mode of propagation is closely related to squirmer models~\cite{light52,Lauga_2009} and droplets driven by Marangoni flows~\cite{schmitt16,Zottl_2016,whit16}, and it can be observed when embedding active surfaces into a passive fluid~\cite{witt23}.

We now want to exploit the capacity of our functional formulation to directly compute nonlinear stationary states in order to characterize the relationship between surface geometry, propagation velocity and key parameters of active surface swimmers in more detail. In particular, we will focus on the interplay of the hydrodynamic length $L_h = \sqrt{\eta_b/\Gamma}$ relative to system size $R_0$, and the P\'eclet number $\text{Pe} = \xi\Delta \mu{R_0^2}/(D\eta_b)$ that characterizes the strength of active contractility~\cite{Bois2011PRL}. To this end, we perform direct computations of stationary geometries and flows on a $500\times500$ grid in the ($\text{Pe}$, $R_0/L_h$) space and evaluate the propagation speed (Fig.~\ref{fig:swimana}a) and a heuristic measure of geometric polarity, $P = 1 - H(u=0)/H(u = 1)$, where $H$ is the mean curvature (Fig.~\ref{fig:swimana}b). This data provides insights into how hydrodynamic screening and activity control surface shape and propagation speed. For $\text{Pe}<\text{Pe}^*$ with Pe$^* = (2+kR_0^2/D)/[\hat{c}_0f'(\hat{c}_0)]$ and $f(c)=c^2/(c^2+c_s^2)$ a spherical surface with homogeneous stress regulator pattern is linearly stable~\cite{Mietke2019Self}. For $\text{Pe}>\text{Pe}^*$ mechanochemical shape instabilities emerge that lead to shape deformations and surface propagation. In the presence of strong hydrodynamic screening ($R_0/L_h\gg1$), it is dominantly the P\'eclet number that affects shape polarity and propagation speed, which both increase with increasing activity (Fig.~\ref{fig:swimana}a,b Pe$^*\approx 22$). This impact of the P\'eclet number was expected, given that it quantifies the strength of contractility and therefore essentially the magnitude of tangential flows that move the surface forward (Fig.~\ref{fig:swimana}c). In the regime $R_0/L_h\approx1$, where hydrodynamic screening starts to directly affect active stress propagation across the surface, changes of the hydrodynamic length have a strong impact on the surface polarity, while still having relatively little impact on the propagation velocity. Exemplary stationary geometries, flows and stress regulator patterns shown in Fig.~\ref{fig:swimana}d suggest an explanation for this phenomenology: Strong hydrodynamic screening for large $R_0/L_h$ leads to stationary stress regulator patterns that put a large part of the rear surface under high tension (see e.g. blue triangle in Fig.~\ref{fig:swimana}d). As a result, the surface is significantly flattened on one side and shape polarity increases. The lack of change in propagation velocities when $R_0/L_h$ is increased appears to be the result of two counter-acting factors: As the area of high stress regulator concentration patch at the rear increases, contractility gradients -- and therefore tangential flows -- localize strongly in the equatorial region ($u\approx1/2$). This leads to a more effective propulsion that compensates for the fact that surface flow magnitudes themselves are strongly reduced when $R_0/L_h$ increases, such that overall surface propagation speeds remain roughly constant. 

\subsubsection{Guided division}\label{sec:SOsurf2}

%%%%%%%%%%%%%%%%%%
%%%%%%%%%%%%%%%%%%
\begin{figure*}[t]
	\includegraphics[width = 2.05\columnwidth]{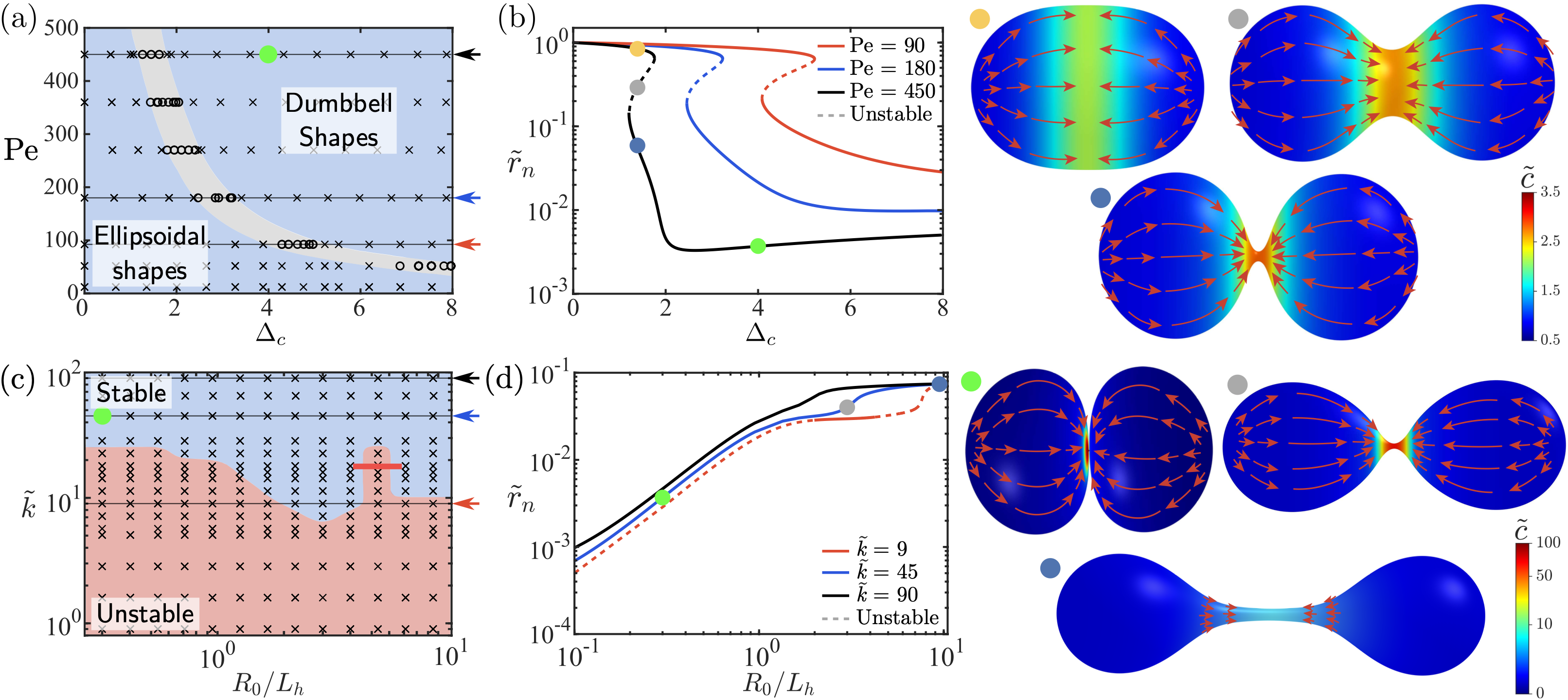}\vspace{-0.2cm}
	\caption{Shape space analysis of guided division model with parameters used in Fig.~\ref{fig:furrowdyn} indicated by green circles.
    \textbf{(a)}~Phase space of stationary surface geometries and flows -- as function of P\'eclet number and magnitude $\Delta_c$ of contractile ring pattern [see Eq.~(\ref{eq:c0dyn})] -- obtained by direct computation of nonlinear stationary surface solutions using the approach described in Sec.~\ref{sec:statsurf}. Black symbols indicate complementary dynamic simulations to test stability of the stationary states. Crosses correspond to regions with unique attractors: dumbbell-shaped and ellipsoidal surfaces. Black circles identify parameter regions with multiple attractors. \textbf{(b)}~1D slices through phase space in (a) show stable (solid lines) and unstable (dashed lines) regions of solution branches and reveals a first-order transition between stationary ellipsoidal and dumbbell shapes. Exemplary stationary solutions in the degenerate part of the shape space [gray region in (a)] are shown on the right. Red arrows indicate surface flows, colour code indicates stress generator concentration $\tilde{c}=c/c_0$. \textbf{(c)}~Stability diagram of symmetrically ingressed stationary surfaces. Black crosses depict parameters for which dynamic simulations were performed to test the mechanical surface stability (red: unstable, blue: stable). The instability ``finger" (passed through by red line) corresponds to a reentrant region in which the symmetric geometry is briefly unstable, while two stable geometries with slightly asymmetrically placed rings emerge (see SI~Fig.~\ref{fig:S1}). \textbf{(d)}~Slices through phase space in (c) show stable (solid lines) and unstable (dashed lines) branches for different turnover rates (Movie~5). Green circles depict parameters used in Fig.~\ref{fig:furrowdyn}. Exemplary stationary solutions along the fully stable branch with $\tilde{k}=45$ are shown on the right. All other parameters are listed in Table~\ref{tab:para_surface}.}
	\label{fig:furrowshsp}
\end{figure*}
%%%%%%%%%%%%%%%%%%
%%%%%%%%%%%%%%%%%%

We finally characterize surface shapes that emerge when turnover of the stress regulator is spatially inhomogeneous [$\Delta_c>0$, see Eq.~(\ref{eq:c0dyn})] with an elevated recruitment rate around the equator ($k_{\text{on}}=k\hat{c}_0$). A representative example of the shape dynamics is depicted in Fig.~\ref{fig:furrowdyn}. We set $c_s=10c_0$ in Eq.~(\ref{eq:stresssc}) to avoid saturating local active tension in regions of high stress regulator concentration, which leads to a critical P\'eclet number Pe$^*\approx2.4\times10^3$. For the analysis, we consider \hbox{Pe $ <500$}, far below the mechanochemical instability threshold. Driven by the turnover profile Eq.~(\ref{eq:c0dyn}), a contractile ring forms and persists if the turnover rate~$k$ is sufficiently large (Fig.~\ref{fig:furrowdyn}). The resulting contractile tension can lead to an almost complete ingression of the surface (neck radius $r_n:=r(u=1/2)$) over time as shown in Fig.~\ref{fig:furrowdyn}a for Pe $=450$. A finite-sized neck remains due to the presence of bending rigidity~$\kappa$. Snapshots of the corresponding dynamics in Fig.~\ref{fig:furrowdyn}b show surface flows towards the contractile ring as previously described in experiment and theory~\cite{bray1988cortical,Mietke2019Minimal,Turlier2014}. The final snapshot reveals a numerically well-resolved, extreme localized curvature ($|R_0C_u^{u}|\approx300$ across $\Delta u\approx10^{-2}$, see inset Fig.~\ref{fig:furrowdyn}b) in the neck region, highlighting the broad geometric regimes our numerical approach can robustly~\hbox{capture}. 

To gain a deeper understanding of the overall space of emerging surface geometries, we use the approach described in Sec.~\ref{sec:dcomp} to directly compute stationary geometries and flows on dense parameter grids, and characterize emergent surface geometries and their stability (Fig.~\ref{fig:furrowshsp}, green circles indicate parameters used in Fig.~\ref{fig:furrowdyn}). Fixing $\tilde{k}=45$ and $R_0/L_h=0.3$, a parameter sweep in the $(\Delta_c$,Pe)-space reveals three regimes~(Fig.~\ref{fig:furrowshsp}a): i)~For small contractility (quantified by the P\'eclet number Pe) or small inhomogeneity magnitude $\Delta_c$, final stationary surfaces are ellipsoidal with positive Gaussian curvature everywhere. ii)~When both, contractility and inhomogeneity magnitude, are sufficiently large dumbbell shapes with a neck (locally negative Gaussian curvature) emerge. iii)~In an intermediate regime, both ellipsoidal and dumbbell-shaped stationary surfaces can be found for fixed Pe and $\Delta_c$ (gray region in Fig.~\ref{fig:furrowshsp}a). Numerical shape perturbations suggest all surfaces are mechanically stable in regimes i) and ii) (blue regions in Fig.~\ref{fig:furrowshsp}a). Tracking the neck radii of stationary geometries (Fig.~\ref{fig:furrowshsp}b) for different fixed P\'eclet numbers (arrows in Fig.~\ref{fig:furrowshsp}a), we find indeed overhangs in the solution branch when crossing between regions of ellipsoidal and dumbbell shapes, indicating a first-order shape transitions between these geometries. Specifically, two stable and one unstable solution (solid and dashed lines Fig.~\ref{fig:furrowshsp}b, respectively) exist in this regime for equal parameters. Exemplary surface geometries for fixed $\Delta_c$ and Pe in this degenerate regime are shown on the right of Fig.~\ref{fig:furrowshsp}b, confirming the first-order transition occurs between surface geometries without (yellow circle) and with (blue circle) neck.

Finally, we want to challenge the stability of the symmetrically ingressed geometries found in Figs.~\ref{fig:furrowshsp}a,b by modifying the turnover rate $k$. Motivated by the observation in Fig.~\ref{fig:swimana}, we additionally investigate if and how hydrodynamic screening imposed by the hydrodynamic length $L_h=\sqrt{\eta_b/\Gamma}$ affects stationary surface geometries. Fixing Pe$\,=450$ and $\Delta_c=4$, a direct computation of symmetrically ingressed stationary geometries and flows in the ($\tilde{k},R_0/L_h)$-space leads to the stability diagram shown in Fig.~\ref{fig:furrowshsp}c. We find that the turnover rate~$k$ has to be larger than a certain threshold for symmetrically ingressed stationary surfaces to be mechanically stable. This can be understood intuitively. We know from the previous analysis for $\Delta_c=0$ (Figs.~\ref{fig:RingSlip},  \ref{fig:swimana}) that self-organized centered contractile rings are generally unstable and contractility patterns exhibit an inherent propensity to polarize the surface. We expect this ring-slipping to occur on some time scale, which the turnover rate $k$ has to compete against when restoring the preferred centered stress-regulator concentration profile given in Eq.~(\ref{eq:c0dyn}). The interface between stable and unstable domains forms a rugged, non-monotonic interface in the ($\smash{\tilde{k},R_0/L_h)}$-space. This interface includes a "finger" region in which symmetrically ingressed stationary surfaces become unstable and stable again only by changing $R_0/L_h$ (red line in Fig.~\ref{fig:furrowshsp}c). A careful analysis of all stationary branches in this regime reveals the occurrence of an additional bifurcation at which the symmetric surface becomes unstable. At this point, two stable branches with a slightly asymmetrically placed ingression appear and merge again (see SI~Fig.~\ref{fig:S1}), indicating a symmetry-breaking and -restoring reentrant behavior when changing the hydrodynamic length~$L_h$.

Shape trajectories in terms of their neck radius for different fixed turnover rates are shown in Fig.~\ref{fig:furrowshsp}d, together with snapshots of stationary surfaces along a fully stable branch and for different hydrodynamic lengths. Analog to the results from Fig.~\ref{fig:MembrShapeSpace}a, changes in hydrodynamic length have the most significant impact around $R_0/L_h\approx 1$, when it changes from being irrelevant to becoming relevant on the surface given the characteristic domain size $\sim R_0$. Emerging geometries change from sharply divided surfaces to more lengthy dumbbell geometries by the same mechanism discovered in Sec.~\ref{sec:SOsurf1}: Hydrodynamic screening for $R_0/L_h>1$ leads to more spread out steady state patterns of the stress regulator. This results in a broader surface region with high contractile tension and consequently the neck region expands (blue circle,~Fig.~\ref{fig:furrowshsp}d). 

\section{Conclusion}
We have introduced a framework that integrates the non-equilibrium physics of deforming active surfaces into an established, highly successful variational approach previously used to analyze shape spaces of equilibrium surfaces. The corresponding dissipation functional is generically constructed from the thermodynamic entropy production and exploits the symmetries imposed by Onsager relations to formulate linear constitutive laws exclusively in terms of dissipative phenomenological coefficients. Such an approach is immediately applicable the large variety of unexplored constitutive laws for active surfaces with different broken symmetries that have recently been derived~\cite{GuillaumePRR2022}. While the restriction to linear couplings may sound restrictive, it is an acceptable limitation given the remarkably broad success of such linear constitutive laws in quantitatively explaining many biological processes~\cite{Prost2015naturephysics}. 

Many aspects of this work benefit from the simplifying restriction to axisymmetric surfaces. It will be interesting to investigate which of these insights can be carried over to surfaces without pre-imposed symmetry. While the unconstrained dissipation functional is valid for arbitrary surfaces, one may ask how the analog of the scaling Lagrangian-Eulerian parameterization could look like on a general surface. This could be guided by the aim to minimize the distortion in the map from a reference space to the deforming surface. While on the axisymmetric surface this corresponds to the elimination of inhomogeneous parametric stretching along the meridional outline, one would also have to minimize the effects of shearing and twisting in the parameterization of a general two-dimensional surface. 

We have considered two minimal models of self-organising active fluid surfaces to exemplify the systematic analysis of stationary shape spaces such models give rise to. The code to perform these analyses, using a standard boundary value solver implemented in \textsc{Matlab}, is available on Github~\cite{git}. It is tempting to use the resulting characterization of stereotypic shape transitions as continuous or discontinuous non-equilibrium phase transitions to speculate about the biological function of either: Discontinuous transitions may be energetically more demanding to achieve, but lead to a new state that is less susceptible for undesired reversals. A complementary point can be made for continuous transitions. Studying alternative constitutive laws in the future therefore promises not only to reveal a rich phenomenology in the dynamics of active surfaces, but also to advance our understanding of the physical principles that underpin shape emergence and its robustness in developing organisms. 

\section{Acknowledgment}
We acknowledge financial support from the National Natural Science Foundation of China under Grant No. 12474199 and the Fundamental Research
Funds for Central Universities of China under Grant No.
20720240144, and 111 project B16029.

% \bibliography{references} 

%

\clearpage
\onecolumngrid

\renewcommand\thefigure{S\arabic{figure}}    
\setcounter{figure}{0} 
\appendix
\section{Differential geometry of axisymmetric surfaces} 
\subsection{Geometric surface properties}
For axisymmetric surfaces described by $\mathbf{X}(u,\phi,t)$ given in Eq.~(\ref{eq:Xaxi}), tangent vectors $\mathbf{e}_i=\partial_i\mathbf{X}$ and surface normal \hbox{$\mathbf{n}=\mathbf{e}_u\times\mathbf{e}_{\phi}/|\mathbf{e}_u\times\mathbf{e}_{\phi}|$} are given by
\begin{equation}\label{eq:bvs}
\mathbf{e}_u=h\begin{pmatrix}
\cos\psi\cos\phi \\
\cos\psi\sin\phi \\
-\sin\psi
\end{pmatrix}
\quad
\mathbf{e}_{\phi}=r\begin{pmatrix}
-\sin\phi\\
\cos\phi \\
0
\end{pmatrix}\\
\quad
\mathbf{n}=\begin{pmatrix}
\sin\psi\cos\phi \\
\sin\psi\sin\phi \\
\cos\psi
\end{pmatrix}.
\end{equation}

The metric tensor $g_{ij}=(\partial_i\mathbf{X})\cdot(\partial_j\mathbf{X})$ and its inverse $g^{ij}=(g_{ij})^{-1}$, as well as the curvature tensor \hbox{$C_{ij}=-\mathbf{n}\cdot\partial_i\partial_j\mathbf{X}$} read
\begin{align}
g_{ij}&=\begin{pmatrix}
g_{uu} & 0\\
0 & g_{\phi\phi}
\end{pmatrix}=\begin{pmatrix}
h^2 & 0\\
0 & r^2
\end{pmatrix}\label{eq:gijaxi}\\
g^{ij}&=
\begin{pmatrix}
h^{-2} & 0\\
0 & r^{-2}
\end{pmatrix}\\
C_{ij}&=\begin{pmatrix}
C_{uu} & 0\\
0 & C_{\phi\phi}
\end{pmatrix}=
\begin{pmatrix}
h\psi'& 0\\
0 & r\sin\psi
\end{pmatrix}\label{eq:Cijaxi},
\end{align}
where we used $h^2=r'^2+z'^2$ (see Eqs.~(\ref{eq:dur}), (\ref{eq:duz}), main text). From the metric tensor we can read off the surface area element
\begin{equation}
dA=\sqrt{g}ds^1ds^2=hrdud\phi.    
\end{equation}

The non-vanishing Christoffel symbols $\Gamma_{ij}^k=\left(\partial_i\partial_j\mathbf{X}\right)\cdot\mathbf{e}^i=\frac{1}{2}g^{kl}\left(\partial_ig_{lj}+\partial_jg_{il}-\partial_lg_{ij}\right)$ are given by
\begin{equation}
\Gamma_{uu}^{u}=\frac{h'}{h}\qquad\Gamma_{u\phi}^{\phi}=\Gamma_{\phi u}^{\phi}=\frac{h\cos\psi}{r}\qquad\Gamma_{\phi\phi}^{u}=-\frac{r\cos\psi}{h}.
\end{equation}

The definition of the covariant derivative 
\begin{equation}\label{eq:defcd}
\nabla_iw_j=\partial_iw_j-\Gamma_{ij}^kw_k    
\end{equation}
implies for a vector field $w_i$ with $\partial_{\phi}w_i=0$ that
\begin{align}
\nabla_uw_u&=w_u'-\frac{w_uh'}{h}\label{eq:duwu}\\
\nabla_{\phi}w_{\phi}&=\frac{r\cos\psi}{h}w_u\label{eq:dpwp}\\
\nabla_uw_{\phi}&=w_{\phi}'-\frac{h\cos\psi}{r}w_{\phi}\label{eq:duwp}\\
\nabla_{\phi}w_u&=-\frac{h\cos\psi}{r}w_{\phi}.\label{eq:dpwu}
% \nabla_uw_u&=w_u'-\Gamma_{uu}^uw_u=w_u'-\frac{w_uh'}{h}\label{eq:duwu}\\
% \nabla_{\phi}w_{\phi}&=-\Gamma_{\phi\phi}^uw_u=\frac{r\cos\psi}{h}w_u\label{eq:dpwp}\\
% \nabla_uw_{\phi}&=w_{\phi}'-\Gamma_{u\phi}^{\phi}w_{\phi}=w_{\phi}'-\frac{h\cos\psi}{r}w_{\phi}\label{eq:duwp}\\
% \nabla_{\phi}w_u&=-\Gamma_{u\phi}^{\phi}w_{\phi}=-\frac{h\cos\psi}{r}w_{\phi}\label{eq:dpwu}
\end{align}
Covariant derivatives of second rank tensors are defined by $\nabla_iT_{jk}=\partial_iT_{jk}-\Gamma^l_{ij}T_{lk}-\Gamma^l_{ik}T_{jl}$ and can be expressed on an axisymmetric surfaces using the expressions above.

\subsection{Vector and tensor divergence}
We provide for reference explicit expressions for the divergences
\begin{equation}
\div(\mathbf{v})=\mathbf{e}^i\cdot\partial_i\mathbf{v}=\nabla_iv^i+C_k^{\,k}v_n\qquad\div(\mathbf{T})=\mathbf{e}^i\cdot\partial_i\mathbf{T}=\nabla_i\mathbf{t}^i  
\end{equation}
of a vector and of a tension tensor field $\mathbf{v}$ and $\mathbf{T}$, respectively, on an axisymmetric surface parameterized by mesh coordinates $u,\phi$, and $h=\sqrt{r'^2+z'^2}$ spatially constant. It is convenient to use normalized tangent basis vectors $\bar{\mathbf{e}}_u:=\mathbf{e}_u/h$ and $\bar{\mathbf{e}}_{\phi}:=\edp/r$, with $\mathbf{e}_i$ given in Eqs.~(\ref{eq:bvs}), to define the components by $\mathbf{v}=\bar{v}_i\bar{\mathbf{e}}_i+v_n\mathbf{n}$ and $\TT = \bar{T}_{ij}\bar{\mathbf{e}}_i\otimes\bar{\mathbf{e}}_j + \bar{T}_{n,i}\bar{\mathbf{e}}_i\otimes\mathbf{n}$. Vector and tensor divergence are then given by
\begin{align}\label{eq:divvexp}
\div({\bf v}) = \frac{1}{hr}\partial_u\left(r\bar{v}_u\right) + \left(\frac{\psi'}{h}+\frac{\sin\psi}{r}\right)v_n
%\frac{\bar{v}_u'}{h}+\frac{\cos\psi}{r}\bar{v}_u + \left(\frac{\psi'}{h}+\frac{\sin\psi}{r}\right)v_n   
\end{align}
and
\begin{align}
    \eu\cdot\div(\TT) &= \frac{1}{hr}\partial_u\left(r\,\bar{T}_{uu}\right)+\frac{\psi'}{h}\bar{T}_{n,u}-\frac{\cos\psi}{r}\bar{T}_{\phi \phi}\label{eq:divTu}\\
    \ep\cdot\div(\TT) &= \frac{1}{hr}\partial_u\left(r\,\bar{T}_{u\phi}\right)+\frac{\sin\psi}{r}\bar{T}_{n,\phi}+\frac{\cos\psi}{r}\bar{T}_{\phi u}\label{eq:divTp}\\
    \mathbf{n}\cdot\div(\TT) &= \frac{1}{hr}\partial_u\left(r\,\bar{T}_{n,u}\right)-\frac{\psi'}{h}\bar{T}_{uu}-\frac{\sin\psi}{r}\bar{T}_{\phi\phi}\label{eq:divTn}.    
\end{align}

\section{Arbitrary Lagrangian-Eulerian (ALE) dynamics of metric and curvature tensors}
Dynamic equation for the metric and curvature tensor of surfaces parameterized by ALE coordinates can be derived in two steps. First, we consider a general shape variation $\mathbf{X}(s^1,s^2)\rightarrow\mathbf{X}(s^1,s^2)+\delta\mathbf{X}(s^1,s^2)$ with $\delta\mathbf{X}=\delta X^i\mathbf{e}_i+\delta X_n\mathbf{n}$. Using the definitions of metric and curvature tensor, as well as $\delta\mathbf{n}=-\left(\mathbf{n}\cdot\partial_i\delta\mathbf{X}\right)\mathbf{e}^i$ and $\nabla_i\mathbf{e}_j=-C_{ij}\mathbf{n}$ one finds~\cite{Guillaume2017}
\begin{align}
\delta g_{ij}&=\nabla_i\delta X_j+\nabla_j\delta X_i+2C_{ij}\delta X_n\label{eq:dgij}\\
\delta C_{ij}&=C^k_{\,j}\nabla_i\delta X_k+C^k_{\,i}\nabla_j\delta X_k +\delta X^k\nabla_kC_{ij}-\nabla_i\nabla_j\delta X_n+C_{ki}C^k_{\,j}\delta X_n.
\end{align}
The Jacobi formula $\delta g=gg^{ij}\delta g_{ij}$ provides an additional relation for the variaton for the metric determinant $g$, which yields together with Eq.~(\ref{eq:dgij})
\begin{equation}
\delta\sqrt{g}=\sqrt{g}\left(\nabla_k\delta X^k+C_k^{\,k}\delta X_n\right).
\end{equation}
In the second step, we assume $\delta\mathbf{X}$ results from an explicit time-dependence as dictated by Eq.~(\ref{eq:ExpGenDyn}). Therefore, we substitute \hbox{$\delta\mathbf{X}=(\mathbf{v}-\mathbf{q})\delta t$} in the variation identities above, where \hbox{$\mathbf{q}=q^i\mathbf{e}_i$} corresponds to a general coordinate flow that depends on the chosen parameterization. This yields the dynamics equations
\begin{align}
\partial_tg_{ij}&=\nabla_i\left(v_j-q_j\right)+\nabla_j\left(v_i-q_i\right)+2C_{ij}v_n\label{eq:dtgij}\\
\partial_tC_{ij}&=C^k_{\,j}\nabla_i(v_k-q_k)+C^k_{\,i}\nabla_j(v_k-q_k) +(v^k-q^k)\nabla_kC_{ij}-\nabla_i\nabla_jv_n+C_{ki}C^k_{\,j}v_n\label{eq:dtCij}\\
\partial_t\sqrt{g}&=\sqrt{g}\left[\nabla_k\left(v^k-q^k\right)+C_k^{\,k}v_n\right]\label{eq:dtg}
\end{align}
for surfaces with an arbitrary Lagrangian-Eulerian parameterization. Known expressions for pure Lagrangian or Eulerian parameterizations~\cite{Guillaume2017} are obtained by setting $q^i=0$ or $q^i=v^i$, respectively.\\

Using Eqs.~(\ref{eq:gijaxi}), (\ref{eq:Cijaxi}), (\ref{eq:duwu}), (\ref{eq:dpwp}) and $w_u=g_{uu}w^u=h^2w^u$, Eq.~(\ref{eq:dtgij}) implies on axisymmetric surfaces the dynamic equations
\begin{align}
\partial_t h & = [(v^u-q^u)h]'+v_n \psi'\label{eq:dtpsiapp}\\
\partial_t r & = (v^u-q^u)h\cos\psi+v_n \sin\psi\label{eq:dtrapp},    
\end{align}
which recovers Eqs.~(\ref{eq:dtr}) and (\ref{eq:dth}) derived in the main text directly by evaluating \linebreak\hbox{$d\mathbf{X}(u,\phi,t)/dt=\mathbf{v}$}. In a similar fashion, we find from the above relations explicit expression for the dynamics of tangent angle $\psi$ and meridional curvature $C^u_{\,u}=\psi'/h$:
\begin{align}
\partial_t \psi &= (v^u-q^u)\psi'-\frac{v_n'}{h}\label{eq:dtpsi}\\
\partial_t C^u_{\,u} &= \left[(v^u-q^u)\frac{\psi'}{h}\right]'-\left(\frac{\psi'}{h}\right)^2v_n-\frac{1}{h}\left[\frac{v_n'}{h}\right]'.\label{eq:dtpsipr}
\end{align}

\section{Conservation laws on ALE-parameterized curved surfaces}
In order to derive conservation laws on ALE-parameterized curved surfaces, one has to understand how quantities of the form
\begin{equation}\label{eq:Forigapp}
F(t)=\int_{\omega(t)}f(s^1,s^2,t)dA
\end{equation}
evolve in time, where the integration domain $\omega(t)$ is a \textit{coordinate region that parameterizes a fixed set of material elements} and $dA=\sqrt{g}\,ds^1ds^2$ is the area element. By definition, the coordinates in $\omega$ will only be independent of time if Lagrangian material coordinates are used. In this case, \smash{$\frac{dF}{dt}$} only has a contribution from temporal changes of the integrand. For~$F(t)$ to be conserved, $\frac{dF}{dt}=0$, one consequently requires
\begin{equation}\label{eq:Lcons}
\text{Lagrangian: }\int_{\omega_0}\partial_t(f\sqrt{g})ds^1ds^2=0.    
\end{equation}

With Eulerian coordinates in Eq.~(\ref{eq:Forigapp}), the integration domain $\Omega(t)$ of coordinates associated with a fixed set of material elements becomes explicitly time-dependent and we expect $\frac{dF}{dt}$ to pick up additional contributions. Instead of deriving these additional contributions, a physical argument is typically made that balances changes within a "fixed" area patch against fluxes across the integration boundary. Specifically, it is postulated that conservation of the quantity $F(t)$ for an Eulerian parameterization implies
\begin{equation}\label{eq:Econs}
\text{Eulerian: }\int_{\omega(t)}\partial_t\left(f\sqrt{g}\right)ds^1ds^2=-\oint_{\partial\omega(t)}fv^i\nu_i ds,
\end{equation}
where $ds$ is the arc length coordinate along the curve bounding the area of integration and $\boldsymbol{\nu}=\nu_i\mathbf{e}^i$ is the in plane unit normal of that curve. In the following, we show using an arbitrary Lagrangian-Eulerian parameterization that both Eqs.~(\ref{eq:Lcons}) and (\ref{eq:Econs}) arise as special cases from a single expression for $\frac{dF}{dt}$, with $F(t)$ given in Eq.~(\ref{eq:Forigapp}), when taking into account temporal changes of both the integrand and the integration domain. 

\subsection{Moving boundary integrals with ALE parameterization}\label{app:movbound}
To determine the total time derivative of $F(t)$ in Eq.~(\ref{eq:Forigapp}) for an ALE-parameterized surface, we have to approximate (denoting for brevity $\tilde{f}=f\sqrt{g}$)
\begin{align}\label{eq:dF}
F(t+\delta t)=\int_{\omega(t+\delta t)}\tilde{f}(\bar{s}^1,\bar{s}^2,t+\delta t)d\bar{s}^1d\bar{s}^2,
\end{align}
to linear order in $\delta t$. The coordinate domains of integration at times $t$ and $t+\delta t$ are parameterized by coordinates~$s^i$ and $\bar{s}^i$, respectively. These coordinates are related by
\begin{align}
\bar{s}^i&=s^i(t+\delta t)\nonumber\\
&=s^i(t)+\left.\frac{\partial s^i}{\partial t}\right|_{S}\delta t+\mathcal{O}(\delta t^2)\nonumber\\
&=s^i+q^i\delta t+\mathcal{O}(\delta t^2),\label{eq:coordt}
\end{align}
where the time derivative at fixed material coordinate $S$ takes into account that we integrate at all times over a fixed set of material elements. The coordinate flow components $q^i=\partial_ts^i|_S$ are in principle arbitrary and may themselves depend on $s^i$. From Eq.~(\ref{eq:coordt}), it then follows that
\begin{equation}\label{eq:fexp}
\tilde{f}(\bar{s}^1,\bar{s}^2,t+\delta t)=\tilde{f}(s^1,s^2,t)+\left(\partial_t\tilde{f}+q^i\partial_i\tilde{f}\right)\delta t+\mathcal{O}(\delta t^2)
\end{equation}
and
\begin{align}
d\bar{s}^1d\bar{s}^2&=\det\left(\frac{\partial\bar{s}^i}{\partial s^j}\right)ds^1ds^2\nonumber\\
&=\left(1+\partial_iq^i\delta t\right)ds^1ds^2+\mathcal{O}(\delta t^2).\label{eq:areael}
\end{align}
Using Eqs.~(\ref{eq:fexp}) and (\ref{eq:areael}) in Eq.~(\ref{eq:dF}) together with Eq.~(\ref{eq:Forigapp}), we find
\begin{equation}
F(t+\delta t)-F(t)=\delta t\int_{\Omega(t)}\left[\partial_t\tilde{f}+\tilde{f}\partial_iq^i+q^i\partial_i\tilde{f}\right]ds^1ds^2+\mathcal{O}(\delta t^2).
\end{equation}
and therefore
\begin{equation}\label{eq:dFdtfin1}
\frac{dF}{dt}=\int_{\omega(t)}\left[\partial_t\tilde{f}+\partial_i(\tilde{f}q^i)\right]ds^1ds^2.
\end{equation}
Reintroducing~$\tilde{f}=f\sqrt{g}$, and using $dA=\sqrt{g}\,ds^1ds^2$ and  $\partial_i\sqrt{g}=\sqrt{g}\,\Gamma_{ik}^k$~\cite{krey68}, Eq.~(\ref{eq:dFdtfin1}) yields
\begin{equation}\label{eq:dtFfin2}
\frac{dF}{dt}=\int_{\omega(t)}\left[\frac{1}{\sqrt{g}}\partial_t(f\sqrt{g})+\nabla_i(fq^i)\right]dA.
\end{equation}
Equation~(\ref{eq:dtFfin2}) holds for any parameterization described by the coordinate flow components~$q^i$. In flat space ($g_{ij}=\delta_{ij}\Rightarrow\sqrt{g}=1,\nabla_i\rightarrow\partial_i$) and with an Eulerian parameterization ($q^i=v^i$) Eq.~(\ref{eq:dtFfin2}) is known as Reynolds transport theorem~\cite{Mandadapu2017}. An alternative way of writing Eq.~(\ref{eq:dtFfin2}) follows from using the metric determinant dynamics Eq.~(\ref{eq:dtg}), which yields
\begin{align}
\frac{dF}{dt}&=\int_{\omega(t)}\left[\partial_tf+q^i\partial_if+f\left(\nabla_kv^k+C_k^{\,k}v_n\right)\right]dA\nonumber\\
&=\int_{\omega(t)}\left[\frac{df}{dt}+f\div\left(\mathbf{v}\right)\right]dA\label{eq:dtFfin3},
\end{align}
where we have used $\div(\mathbf{v}):=\mathbf{e}^i\cdot\partial_i\mathbf{v}=\nabla_kv^k+C_k^{\,k}v_n$ and the definition of the total time derivative $\frac{d}{dt}(\cdot)$ for arbitrary Lagrangian-Eulerian coordinates [see also Eq.~(\ref{eq:ExpGenDyn})]
\begin{equation}
\frac{df}{dt}:=\partial_tf+q^i\partial_if.    
\end{equation}

Note, that using Eq.~(\ref{eq:dtFfin3}) with $f=1$ the dynamics of the total surface area follows as 
\begin{equation}
\frac{dA}{dt}=\int\left(\nabla_kv^k+C_k^{\,k}v_n\right)dA=\int\div\left(\mathbf{v}\right)dA,
\end{equation}
which is independent of the choice of the coordinate flow components $q^i$. Therefore, local area conservation on the surface can be realized by imposing $\div(\mathbf{v})=0$, irrespective of the choice of parameterization and as done in Sec.~\ref{sec:membr} via the Lagrange multiplier $\gamma_H$ [see Eq.~(\ref{eq:dFdtmembr})]. The explicit expression of $\div\left(\mathbf{v}\right)$ on the cylindrical surface is given in Eq.~(\ref{eq:divvexp}). 

\subsection{Generalized conservation laws and continuity equation}
If $F(t)$ is a conserved quantity, we must have $\frac{dF}{dt}=0$ and the right-hand side of Eq.~(\ref{eq:dtFfin2}) can be rearranged into
\begin{equation}\label{eq:Acons}
\text{ALE: }\int_{\Omega(t)}\partial_t\left(f\sqrt{g}\right)ds^1ds^2=-\oint_{\partial\Omega(t)}fq^i\nu_i ds,    
\end{equation}
where we have used the covariant Stokes theorem~\cite{Guillaume2017}
\begin{equation}\label{eq:ST}
\int_{\Omega}\nabla_iw^idA=\oint_{\partial\Omega}w^i\nu_i ds.   
\end{equation}
%Equation~(\ref{eq:Acons}) forms the basis to derive conservation laws using arbitrary parameterizations. 
For the special cases of Lagrangian material coordinates ($q^i=0$) and Eulerian coordinates ($q^i=v^i$), we recover from the general expression (\ref{eq:Acons}) the known expressions Eqs.~(\ref{eq:Lcons}) and~(\ref{eq:Econs}). For any other parameterization choice, one still finds an effective "flux across the boundary" $\sim f\mathbf{q}$, which is however just an artefact of the parameterization and has no immediate physical interpretation.

\subsubsection{Mass conservation}\label{app:masscons}
Using Eq.~(\ref{eq:dtFfin3}) with $f=\rho$, where $\rho$ is the mass area density, and imposing that total mass $F(t)=M(t)$ must be conserved, we find
\begin{equation}\label{eq:mcgen}
\partial_t\rho+q^i\partial_i\rho+\rho\,\div(\mathbf{v})=0.    
\end{equation}
It is easy to check that mass conservation law Eq.~(\ref{eq:mcgen}) takes the known forms when considering purely Lagrangian ($q^i=0$) or purely Eulerian coordinates ($q^i=v^i$)~\cite{Guillaume2017}. On an axisymmetric surface, we can use Eq.~(\ref{eq:divvexp}) to express Eq.~(\ref{eq:mcgen}) as
\begin{equation}\label{eq:dtrhoaxi}
\partial_t\rho+q^u\rho'+\frac{\rho}{hr}\left[\partial_u\left(r\bar{v}_u\right) + \left(r\psi'+h\sin\psi\right)v_n\right]=0.
\end{equation}
Equation (\ref{eq:dtrhoaxi}) is used in Secs.~\ref{sec:SOsurf1} and \ref{sec:SOsurf2} to evolve a concentration field on the surface.

\subsubsection{Momentum conservation and force balance}\label{app:momcons}
The total linear momentum of an arbitrarily parameterized patch of fixed material elements is given by
\begin{equation}\label{eq:momentum}
\mathbf{g}=\int_{\Omega(t)}\rho\mathbf{v}dA. 
\end{equation}
Momentum balance requires
\begin{align}\label{eq:mombalgen}
\frac{d\mathbf{g}}{dt}=\oint_{\partial\Omega}\boldsymbol{\nu}\cdot\mathbf{T}ds+\int_{\Omega}\mathbf{f}^{\text{ext}}dA,
\end{align}
where $\mathbf{T}=\mathbf{e}_i\otimes\mathbf{t}^i$ is the tension tensor and $\mathbf{f}^{\text{ext}}$ are external forces. Equation~(\ref{eq:mombalgen}) must hold for any fixed set of material elements. Hence, combining Eq.~(\ref{eq:dtFfin3}), mass conservation Eq.~(\ref{eq:mcgen}) and the covariant Stokes theorem Eq.~(\ref{eq:ST}), momentum balance can be expressed as
\begin{align}
\rho\mathbf{a}=\nabla_i\mathbf{t}^i+\mathbf{f}^{\text{ext}},
\end{align}
where $\nabla_i\mathbf{t}^i=\div(\mathbf{T})$ and
\begin{equation}
\mathbf{a}:=\frac{d\mathbf{v}}{dt}=\partial_t\mathbf{v}+q^i\partial_i\mathbf{v}    
\end{equation}
is the local center-of-mass acceleration for an arbitrary Lagrangian-Eulerian parameterization. Explicit component expressions of $\nabla_i\mathbf{t}^i$ on axisymmetric surfaces, most relevant for the overdamped force balance considered in this work, are provided in Eqs.~(\ref{eq:divTu})--(\ref{eq:divTn}). Alternatively, we can use Eq.~(\ref{eq:dtFfin2}) to evaluate time derivative $\frac{d\mathbf{g}}{dt}$ with $\mathbf{g}$ given in Eq.~(\ref{eq:momentum}) and find that momentum balance Eq.~(\ref{eq:mombalgen}) can be equivalently written as
\begin{equation}\label{eq:mombalgen2}
\frac{1}{\sqrt{g}}\partial_t\left(\rho\mathbf{v}\sqrt{g}\right)=\div\left(\mathbf{T}-\rho\,\mathbf{q}\otimes\mathbf{v}\right)+\mathbf{f}^{\text{ext}}.
\end{equation}
In the special case of an Eulerian parameterization, $q^i=v^i$, Eqs.~(\ref{eq:mombalgen})--(\ref{eq:mombalgen2}) are equivalent to expressions in~\cite{Guillaume2017} and, in flat space ($g_{ij}=\delta_{ij}\Rightarrow\sqrt{g}=1,\nabla_i\rightarrow\partial_i$), $\mathbf{T}_{\text{tot}}=\mathbf{T}-\rho\,\mathbf{v}\otimes\mathbf{v}$ is sometimes referred to as "total stress"~\cite{Julicher_2018}. 

\subsubsection{Torque balance}
For completeness, we provide component expression of the torque balance equation on axisymmetric surfaces. Specifically, Eq.~(\ref{eq:TB}) with $\mathbf{M}=\mathbf{e}_i\otimes\mathbf{m}^i$ can be written as
\begin{equation}\label{eq:TBapp}
\nabla_i\mathbf{m}^i=\mathbf{t}^i\times\mathbf{e}_i.    
\end{equation}
The component expressions of $\nabla_i\mathbf{m}^i$ can be read off Eqs.~(\ref{eq:divTu})--(\ref{eq:divTn}), with (normalized) in-plane tension and normal force components replaced by the corresponding moment tensor components, $\bar{t}_{ij}\rightarrow\bar{m}_{ij}$ and $\bar{t}^i_n\rightarrow\bar{m}^i_n$. The vector on the right-hand side of Eq.~(\ref{eq:TBapp}) has components
\begin{align}
    \eu\cdot(\mathbf{t}^i\times\mathbf{e}_i) &= -\bar{T}_{n,\phi}\label{eq:CPu}\\
    \ep\cdot(\mathbf{t}^i\times\mathbf{e}_i) &= \bar{T}_{n,u}\label{eq:CPp}\\
    \mathbf{n}\cdot(\mathbf{t}^i\times\mathbf{e}_i) &= \bar{T}_{\phi u}-\bar{T}_{u\phi},
\end{align}
where we have used~\cite{Guillaume2017}
\begin{equation}
\epsilon_{ij}:=\mathbf{n}\cdot(\mathbf{e}_i\times\mathbf{e}_j)=\sqrt{g}\begin{pmatrix}
0 & 1\\
-1 & 0
\end{pmatrix}
\end{equation}
and $\sqrt{g}=hr$ on the axisymmetric surface. 

\subsubsection{Chemical free energy dynamics}
The free energy density in the rest frame has a contribution from the number density 
\begin{equation}
    f_c = f_c(c^{\alpha}).
\end{equation}
The free energy change rate
\begin{align}
    \frac{dF_c}{dt} & = \int\left(\frac{d f_c}{d t} + \div(\mathbf{v})f_c\right)dA\nonumber\\
    & = \int\left( \mu^{\alpha}\frac{d c^{\alpha}}{dt}+\div(\mathbf{v})f_c\right)dA\nonumber\\
    & = \int\left[ (f_c-\mu^{\alpha}c^{\alpha})\div(\mathbf{v})-\mu^{\alpha}\div(\mathbf{j}^{\alpha}) + (r_c^{\alpha}+J_n^{\alpha})\mu^{\alpha} \right  ] dA,\label{eq:Tebare}
\end{align}
where $\mu^{\alpha} = \partial f_c/\partial c^{\alpha}$ is the chemical potential associated with the $\alpha$ species. When we perform variations against the velocity~$\mathbf{v}$, Eq.~(\ref{eq:Tebare}) will contribute an equilibrium isotropic stress 
\begin{equation}\label{eq:strequi1}
    {\mathbf{T}_{e,0}} = (f_c - \mu^{\alpha}c^{\alpha})\mathbf{G} \equiv \gamma_e\mathbf{G}.
\end{equation}

\section{Equilibrium tension and moments}\label{app:equitens}
Equilibrium tensions and moments arising from the Helfrich free energy $F_H=\int dA f_H$ with $f_H$ given in Eq.~(\ref{eq:helfrich}) (main text) can alternatively be derived using the virtual work principle, which leads for arbitrary surface to~\cite{Guillaume2017}
\begin{align}
t^{ij}_e&=\frac{\kappa}{2}\left(C_k^{\,k}-C_0\right)\left(\left(C_k^{\,k}-C_0\right)g^{ij}-2C^{ij}\right)\label{eq:te}\\
m^{ij}_e&=\kappa\left(C_k^{\ k}-C_0\right)\epsilon^{ij}.\label{eq:me}
\end{align}
The normal component of the torque balance Eq.~(\ref{eq:TBapp}) implies in this case that $t^{ij}_e$ is symmetric. The in-plane component of Eq.~(\ref{eq:TBapp}) yields normal forces of the form~\cite{Guillaume2017}
\begin{equation}\label{eq:tne}
t^i_{n,e}=\kappa\nabla^i\left(C_k^{\,k}-C_0\right).
\end{equation}
For $C_0,\kappa$ constant and $\TT = t^{ij}_e\mathbf{e}_i\otimes\mathbf{e}_j + t^i_{n,e}\mathbf{e}_i\otimes\mathbf{n}$ with tension and normal forces given in Eqs.~(\ref{eq:te}) and (\ref{eq:tne}) the tangential force balance $\smash{\mathbf{e}^j\cdot\div(\mathbf{T}_e)=\nabla_it^{ij}_e+C_i^{\,j}t_{n,e}^i=0}$ is identically satisfied. An inhomogeneous spontaneous curvature, as considered in Sec.~\ref{sec:membr}, will in general contribute a force density to the tangential force balance [see Eq.~(\ref{eq:divTeu})]. On an axisymmetric surface, the Helfrich free energy density reads
\begin{equation}\label{eq:HelfAxi}
f_{\kappa}=\frac{\kappa}{2}\left(\frac{\sin\psi}{r}+\frac{\psi'}{h}-C_0\right)^2.
\end{equation}
Together with Eq.~(\ref{eq:strequi1}) the full equilibrium tensor can then be expressed as
\begin{align}\label{eq:HelfTens}
{\bf T}_e & = \gamma\mathbf{G} + \frac{\kappa}{2}\left[\left(\frac{\sin\psi}{r}-C_0\right)^2-\left(\frac{\psi'}{h}\right)^2\right]\bar{{\bf e}}_u\otimes\bar{{\bf e}}_u + \frac{\kappa}{2}\left[\left(\frac{\psi'}{h}-C_0\right)^2-\left(\frac{\sin\psi}{r}\right)^2\right]\bar{{\bf e}}_{\phi}\otimes\bar{{\bf e}}_{\phi} \nonumber\\
&+\kappa\left[\frac{1}{h}\partial_u \left(\frac{\sin\psi}{r}+\frac{\psi'}{h}-C_0\right)\right] \bar{{\bf e}}_u\otimes {\bf n},
\end{align}
where $\gamma=\gamma_e+\gamma_H$. This implicitly defines the equilibrium tension tensor components $(\bar{T}_e)_{uu}$, ($\bar{T}_e)_{\phi\phi}$ and $(\bar{T}_e)_u^{n}$ with respect to the normalized basis via
\begin{equation}\label{eq:HTcomps}
{\bf T}_e=(\bar{T}_e)_{uu}\bar{\mathbf{e}}_u\otimes\bar{\mathbf{e}}_u+(\bar{T}_e)_{\phi\phi}\bar{\mathbf{e}}_\phi\otimes\bar{\mathbf{e}}_\phi+(\bar{T}_e)_u^{n} \bar{{\bf e}}_u\otimes {\bf n}.
\end{equation}
From Eqs.~(\ref{eq:divTu})--(\ref{eq:divTn}) and (\ref{eq:HelfTens}) it follows that 
\begin{align}
    \div({\bf T}_e)\cdot\ebar_u & = \frac{\gamma'}{h}  + \frac{\partial \fk}{\partial C_0}\frac{C_0'}{h} \label{eq:divTeu}\\
    \div({\bf T}_e)\cdot\ep & = 0,  \label{eq:divTep}
\end{align}
as well as
\begin{align}
\div({\bf T}_e)\cdot\mathbf{n} &= - \left(\frac{\kappa C_0^2}{2}+\gamma\right)\left(\frac{\sin\psi}{r}+\frac{\psi'}{h}\right) + 2\kappa C_0 \frac{\sin\psi}{r} \frac{\psi'}{h } + C_0'\left(\frac{\kappa h'}{h^3}- \frac{\kappa \cos\psi}{h r} \right)- \frac{\kappa C_0''}{h^2}\nonumber\\
&+\frac{\kappa \sin\psi}{4 r^3}[3+\cos(2\psi)] - \frac{\kappa \psi'}{4 h r^2}[1+3\cos(2\psi)]- \frac{2\kappa \cos\psi\, h' \psi'}{h^3 r}
+ \frac{3\kappa (h')^2 \psi'}{h^5}- \frac{\kappa  h'' \psi'}{h^4}\nonumber\\
&- \frac{3\kappa \sin\psi\, (\psi')^2}{2 h^2 r}
+ \frac{\kappa (\psi')^3}{2 h^3}+ \frac{2\kappa \cos\psi\, \psi''}{h^2 r}
- \frac{3\kappa h' \psi''}{h^4}+ \frac{\kappa \psi'''}{h^3},\label{eq:divTen}
\end{align}
which corresponds to the shape equation of axisymmetric Helfrich membranes and agrees for $C_0'=0$ with Eq.~(13) in ref.~\cite{julicher1994}, where the tangent angle was parameterizied as $\psi(r)$.

\section{Equivalence between variational equations and force balance equations}\label{app:FBequi}
We will prove in App.~\ref{app:FBequi} that variational equations derived from the dissipation functional constructed in the main text are equivalent to the force and moment balance equation of the fluid introduced in Sec.~\ref{sec:ActSurf}. To prepare this computation, we first provide explicit expressions of the relevant tensors on axisymmetric surfaces. The isotropic part $\smash{\isoS}$ and the traceless symmetric part $\smash{\tbS}$ [see Eq.~(\ref{eq:defS})] of the strain rate tensor $\smash{\bS}=\frac{1}{2}(\mathbf{e}_i\cdot\partial_j\mathbf{v}+\mathbf{e}_j\cdot\partial_i\mathbf{v})\mathbf{e}^i\otimes\mathbf{e}^j$ can be expressed on axisymmetric surfaces as
 \begin{align}\label{eq:prepid1}
\isoS&=\frac{1}{2}\left[\frac{\bar{v}_u'}{h}+\frac{\cos\psi}{r}\bar{v}_u+\left(\frac{\psi'}{h}+\frac{\sin\psi}{r}\right)v_n\right]\mathbf{G}\\
\tbS&=\frac{1}{2}\left(\frac{\bar{v}_u'}{h}-\frac{\cos\psi}{r}\bar{v}_u\right)\boldsymbol{\tau}_1+\frac{1}{2}\left(\frac{\bar{v}_{\phi}'}{h}-\frac{\cos\psi}{r}\bar{v}_\phi\right)\boldsymbol{\tau}_2+v_n\tilde{\mathbf{C}},\label{eq:prepid2}
 \end{align}
where
\begin{align}
\mathbf{G}&=\left(\bar{\mathbf{e}}_u\otimes\bar{\mathbf{e}}_u+\bar{\mathbf{e}}_\phi\otimes\bar{\mathbf{e}}_\phi\right)\label{eq:prepid3}\\
\boldsymbol{\tau}_1&=\left(\bar{\mathbf{e}}_u\otimes\bar{\mathbf{e}}_u-\bar{\mathbf{e}}_\phi\otimes\bar{\mathbf{e}}_\phi\right)\label{eq:prepid4}\\
\boldsymbol{\tau}_2&=\left(\bar{\mathbf{e}}_u\otimes\bar{\mathbf{e}}_\phi+\bar{\mathbf{e}}_\phi\otimes\bar{\mathbf{e}}_u\right)\label{eq:prepid5}
\end{align}
and
\begin{equation}
\tilde{\mathbf{C}}=\frac{1}{2}\left(\frac{\psi'}{h}-\frac{\sin\psi}{r}\right)\boldsymbol{\tau}_1\label{eq:prepid6}
\end{equation}
is the traceless component of the curvature tensor. From this, we can read the strain rate tensor components with respect to the normalized basis as
 \begin{align}
\bar{S}_{uu} & = \frac{1}{h}\bar{v}_u'+\frac{\psi'}{h}v_n \label{eq:Suu}\\
\bar{S}_{\phi\phi} & = \frac{\cos\psi}{r}\bar{v}_u+\frac{\sin\psi}{r}v_n \\
\bar{S}_{u\phi} & = \frac{1}{2}\left(\frac{\bar{v}_{\phi}'}{h}-\frac{\cos\psi}{r}\bar{v}_{\phi}\right)\label{eq:Sup}.
 \end{align}

Additionally, we introduce a component notation for the dissipative tension tensor $\mathbf{T}_d$ given in Eq.~(\ref{eq:stressd}) (main text),
\begin{equation}
\mathbf{T}_d(\mathbf{v},\Delta\mu)=(\bar{T}_d)_{uu}\bar{\mathbf{e}}_{u}\otimes\bar{\mathbf{e}}_{u}+(\bar{T}_d)_{\phi\phi}\bar{\mathbf{e}}_{\phi}\otimes\bar{\mathbf{e}}_{\phi}+(\bar{T}_d)_{\phi u}\bar{\mathbf{e}}_{\phi}\otimes\bar{\mathbf{e}}_u+(\bar{T}_d)_{u\phi}\bar{\mathbf{e}}_u\otimes\bar{\mathbf{e}}_{\phi},
\end{equation}
where
\begin{align}
(\bar{T}_d)_{uu} & = \eta_b(\bvuu+\bvpp)+\eta_s(\bvuu-\bvpp)+\xi\Delta\mu\label{eq:Tuu}\\
(\bar{T}_d)_{\phi\phi} & = \eta_b(\bvuu+\bvpp)-\eta_s(\bvuu-\bvpp)+\xi\Delta\mu\label{eq:Tpp}\\
(\bar{T}_d)_{u\phi} & =(\bar{T}_d)_{\phi u} = 2\eta_s \bar{S}_{u\phi}\label{eq:Tup}.
 \end{align}

\subsection{Dissipation functional density on an axsymmetric surface}
The density of the Rayleigh dissipation functional in the $(\mathbf{v},r_p)$ ensemble defined by Eq.~(\ref{eq:rayleighaxi}) in the main text reads
\begin{align}
R(\mathbf{v},r_p) & = \left[\partial_t\fk+\fk\frac{\partial_t(r h)}{rh} - r_p \Delta\mu + J_n^{\mathrm{ext}}-\mathbf{f}^{\text{ext}}\cdot\mathbf{v}+\gamma\div(\mathbf{v})+\frac{1}{2}(S_1+S_2+S_3)+\frac{1}{2}\Gamma\mathbf{v}_{\parallel}^2\right]rh, \label{eq:rayldens}
\end{align}
where $\fk$ is given in Eq.~(\ref{eq:HelfAxi}), $\gamma = \gamma_H +\gamma_e$, and
\begin{align}
S_1 & = \eta_b (\bvuu+\bvpp)^2, \\
S_2 & = \eta_s[(\bvuu-\bvpp)^2+(2\bvup)^2], \\
S_3 & = \frac{1}{\Lambda}\left[r_p+\xi(\bvuu+\bvpp) \right]^2
\end{align}
contribute to the density $\hat{\theta}_{\text{int}}= S_1 + S_2 + S_3$ of the internal entropy $\hat{\Theta}_{\text{int}}$ given in Eq.~(\ref{eq:ephat}) (main text). In Eq.~(\ref{eq:rayldens}), we have for generality included any velocity independent external forces (for simplicity using the same symbol $\mathbf{f}^{\text{ext}}$). The time derivative of the bending energy density $\fk$ in Eq.~(\ref{eq:rayldens}) expands into
\begin{equation}\label{eq:dtfk}
    \partial_t \fk = \frac{\partial \fk}{\partial r}\partial_t r + \frac{\partial \fk}{\partial \psi} \partial_t\psi + \frac{\partial \fk}{\partial \psi'} \partial_t \psi' + \frac{\partial \fk}{\partial h} \partial_t h + \frac{\partial \fk}{\partial C_0}\partial_t C_0.
\end{equation}

Finally, we note that variation $\frac{\delta\mathcal{R}}{\delta r_p}=0$ yields the condition $\frac{\partial R}{\partial r_p}=0$, which gives
 \begin{equation}\label{eq:rpconst}
     \frac{1}{\Lambda}\left[r_p+\xi(\bvuu+\bvpp) \right]-\Delta \mu = 0,
 \end{equation}
from which the desired constitutive law for the production rate $r_p$ (Eq.~(\ref{eq:constrp}), main text) follows.

\subsection{Relevance of different parameterization choices for the variation}\label{app:relpara}
When discussing variations in terms of a conventional Euler parameterization ($q^i=v^i$), we substitute the time derivatives in Eqs.~(\ref{eq:rayldens}) and (\ref{eq:dtfk}) by 
\begin{align}
\partial_t r &= v_n \sin\psi\label{eq:dtrvn}\\
\partial_t z &= v_n \cos\psi\\
\partial_t h &= \psi' v_n\\
\partial_t \psi &= -h^{-1}v_n'\\
\partial_t \psi' &= h^{-2}h'v_n' -h^{-1}v_n''\label{eq:dtpspvn}\\
\partial_t C_0&=-h^{-1}\bvu C'_0,\label{eq:dtC0vn}
\end{align}
which follows from Eqs.~(\ref{eq:dtpsiapp})--(\ref{eq:dtpsipr}) and Eq.~(\ref{eq:dC0dt}) (main text). When supplementing the functional density $R(\mathbf{v},r_p)$ in  Eq.~(\ref{eq:rayldens}) by Lagrange multiplier terms introduced in Eq.~(\ref{eq:Lmultiplier}) (main text) to impose geometric relations and SLE parameterization, the time derivatives on the left-hand side of Eqs.~(\ref{eq:dtrvn})--(\ref{eq:dtpspvn}) all become independent variables and
\begin{equation}
\partial_t C_0 = -(\bvu + \partial_t z\sin\psi - \partial_t r \cos\psi)\frac{C_0'}{h},
\end{equation}
which follows from Eqs.~(\ref{eq:quryv}) and (\ref{eq:dC0dt}) in the main text. This implies that -- for the purpose of functional variations with respect to tangential flow velocity components $\bar{v}_u$ and $\bar{v}_\phi$  -- Euler and SLE parameterizations lead immediately to the same Euler-Lagrange equations that amount to the tangential force balance equations. However, variations leading to the normal force balance equation are distinct for Euler and SLE parameterizations (see below).

\subsection{Tangential force balance from variational equations}\label{app:vareultang}
The Euler-Lagrange equation for variations with respect to meridional flows,
\begin{equation}
\frac{1}{2\pi}\frac{\delta \mathcal{R}}{\delta \bvu}=\frac{\partial R}{\partial \bar{v}_u}-\left[\frac{\partial R}{\partial \bar{v}_u'}\right]'=0,
\end{equation}
implies for $R(\mathbf{v},r_p)$ given in  Eq.~(\ref{eq:rayldens})
\begin{equation}
\frac{1}{h}\left[\eta_b(\bvuu+\bvpp)+\eta_s(\bvuu-\bvpp)+\gamma+\xi\Delta\mu \right]' + 2\eta_s(\bvuu-\bvpp)\frac{\cos\psi}{r}-\Gamma \bar{v}_u +\fu  + \frac{\partial \fk}{\partial C_0}\frac{C_0'}{h} = 0,\label{eq:varivu}    
\end{equation}
where we have used Eq.~(\ref{eq:rpconst}) to express $r_p$ in terms of $\Delta\mu$ and $\partial_ur=h\cos\psi$. We see from Eqs.~(\ref{eq:Tuu}) and (\ref{eq:Tpp}) that Eq.~(\ref{eq:varivu}) can be re-written as
\begin{equation}\label{eq:compuvar}
\frac{[\btduu]'}{h}+\frac{\cos\psi}{r}[\btduu-\btdpp] = \Gamma \bar{v}_u-f^{\text{ext}}_u - \frac{\gamma'}{h}-\frac{\partial \fk}{\partial C_0}\frac{C_0'}{h}.
\end{equation}
From Eqs.~(\ref{eq:divTu}) and Eqs.~(\ref{eq:divTeu}) it then follows that Eq.~(\ref{eq:compuvar}) corresponds to the meridional force balance 
\begin{equation}
\bar{{\bf e}}_u \cdot {\rm div}(\mathbf{T}_d +\mathbf{T}_e)=\Gamma\bar{v}_u-f^\text{ext}_u.    
\end{equation}

The Euler-Lagrange equation for variations with respect to azimuthal flows, 
\begin{align}
\frac{1}{2\pi}\frac{\delta \mathcal{R}}{\delta \bvp}&=\frac{\partial R}{\partial \bar{v}_{\phi}}+\left[\frac{\partial R}{\partial \bar{v}_{\phi}'}\right]'=0,
\end{align}
implies
\begin{equation}
\frac{2 \eta}{h} (\bvup)' + 4\eta \bvup \frac{\cos\psi}{r}= \Gamma \bar{v}_{\phi} -\fp. \label{eq:vphi}   
\end{equation}
Comparison with Eq.~(\ref{eq:Tup}) shows that we can rewrite Eq.~(\ref{eq:vphi}) as
\begin{equation}
\frac{\btdup'}{h}+2\frac{\cos\psi}{r}\btdup=\Gamma v_{\phi}-\fp,
\end{equation}
which leads together with Eqs.~(\ref{eq:divTp}) and (\ref{eq:divTep}) to the azimuthal component of the tangential force balance
\begin{equation}
\bar{{\bf e}}_\phi \cdot {\rm div}(\mathbf{T}_d +\mathbf{T}_e)=\Gamma\bar{v}_\phi-f^\text{ext}_\phi.    
\end{equation}

\subsection{Normal force balance from variational equation}
The variational steps that recover the normal force balance equation are distinct for Eulerian parameterization and the SLE parameterization (see Sec.~\ref{app:relpara}) and are therefore discussed separately in the following two sections.

\subsubsection{Unconstrained functional with Eulerian parameterization}\label{app:vareul}.
Using Eqs.~(\ref{eq:dtrvn})--(\ref{eq:dtC0vn}) in Eq.~(\ref{eq:dtfk}), we find
\begin{align}
    \frac{\partial (\partial_t\fk)}{\partial v_n} & = - \kappa\left[\left(\Cpp \right)^2 + \left(\Cuu \right)^2\right]\left(\Cpp + \Cuu - C_0 \right)\\
    \frac{\partial (\partial_t\fk)}{\partial v_n'} & = -\kappa\left(\Cuu+\Cpp-C_0\right)\left(\frac{\cos\psi}{r}-\frac{h'}{h^2}\right)\frac{1}{h}\\
    \frac{\partial (\partial_t\fk)}{\partial v_n''} & -\kappa\left(\Cuu+\Cpp-C_0\right)\frac{1}{h^2},
\end{align}
which can be used to show that the Rayleigh functional density $R(\mathbf{v},r_p)$ in Eq.~(\ref{eq:rayldens}) gives
\begin{align}
\frac{\partial R}{\partial v_{n}} &= \frac{\psi'}{h}\left[\bteuu+\btduu \right] + \frac{\sin\psi}{r}\left[\btepp + \btdpp \right] , \label{eq:dRdvn}\\
\frac{\partial R}{\partial v_{n}'}-\frac{\rm d}{{\rm d}u}\left[\frac{\partial R}{\partial v_{n}''}\right] &= -r\bteun. \label{eq:dRddvn}
\end{align}
Here, we have used Eq.~(\ref{eq:rpconst}) and compared the resulting expressions with the tension tensor components defined by Eqs.~(\ref{eq:HelfTens}) and (\ref{eq:HTcomps}) ($\mathbf{T}_e$), and in Eqs.~(\ref{eq:Tuu}) and (\ref{eq:Tpp}) ($\mathbf{T}_d$). From Eq.~(\ref{eq:divTn}), we see that Eqs.~(\ref{eq:dRdvn}) and (\ref{eq:dRddvn}) together imply
\begin{equation}
    \frac{1}{2\pi rh}\frac{\delta\mathcal{R}}{\delta v_{n}}  =\frac{1}{rh}\left( \frac{\partial R}{\partial v_{n}}-\frac{\rm d}{{\rm d}u}\left[\frac{\partial R}{\partial v_{n}'}\right]+ \frac{\rm d^2}{{\rm d}u^2}\left[\frac{\partial R}{\partial v_{n}''}\right]\right)
    = -{\bf n} \cdot {\rm div}(\hat{\bf{T}}_d+{\bf T}_e)  = 0,
\end{equation}
which is the force balance equation in the normal direction. 

\subsubsection{Constrained functional with SLE parameterization}\label{app:varsle}
A key result of this work is the constrained dissipation functional Eq.~(\ref{eq:Rbar}), which complements the functional density $R(\mathbf{v},r_p)$ given in Eq.~(\ref{eq:rayldens}) by  Lagrangian multiplier terms that impose nonlinear geometric relations and an SLE parameterization. The modified functional density reads [see Eq.~(\ref{eq:Rbar})]
\begin{equation}\label{eq:Rbaraxi}
\bar{R}(\bar{v}_u,\bar{v}_\phi,\partial_t r,\partial_t z, \partial_t h, \partial_t \psi, \partial_t\alpha, \partial_t\beta, \partial_t \zeta,r_p)=R(\mathbf{v},r_p)+\partial_t\left[\alpha(r'-h \cos\psi)+\beta(z'+h\sin\psi)+\zeta h'\right].
\end{equation}

To show how the normal force balance arises from variations of $\bar{R}$ given in Eq.~(\ref{eq:Rbaraxi}), we first note that variations with respect to $\partial_tr$ lead to the Euler-Lagrange equation
\begin{equation}
\frac{\partial\bar{R}}{\partial (\partial_t r)}-\left[\frac{\partial\bar{R}}{\partial (\partial_t r')}\right]'=0,    
\end{equation}
which can be rearranged into
\begin{equation}
\label{eq:alphaprime}
\alpha' = \left[\btduu\frac{\psi'}{h}+\btdpp\frac{\sin\psi}{r} +\gamma\left(\frac{\psi'}{h}+\frac{\sin\psi}{r}\right) \right]rh\sin\psi + (\btepp-\gamma)h-\fn rh \sin\psi + \frac{\partial \fk}{\partial C_0}rC_0'\cos\psi
\end{equation}
and effectively corresponds $r$-component to the normal force balance. Similarly, variations with respect to $\partial_t z$ lead to 
\begin{equation}
\frac{\partial\bar{R}}{\partial (\partial_t z)}-\left[\frac{\partial\bar{R}}{\partial (\partial_t z')}\right]'=0,
\end{equation}
which can be rearranged into
\begin{equation}
\label{eq:betaprime}
\beta' = \left[(\bar{T}_d)_{uu}\frac{\psi'}{h}+(\bar{T}_d)_{\phi\phi}\frac{\sin\psi}{r} + \gamma\left(\frac{\psi'}{h}+\frac{\sin\psi}{r}\right) \right]rh\cos\psi-\fn rh \cos\psi -  \frac{\partial \fk}{\partial C_0}rC_0'\sin\psi
\end{equation}
and effectively corresponds to the $z$-component of the normal force balance. From variations with respect to $\partial_t \psi$ and $\partial_t h$, we find
\begin{align}
\frac{\partial\bar{R}}{\partial (\partial_t \psi)}-\left[\frac{\partial\bar{R}}{\partial (\partial_t \psi')}\right]' &= \frac{\partial \fk}{\partial \psi} r h +\alpha h \sin\psi +\beta h\cos\psi-\left[\frac{\partial \fk}{\partial \psi'} r h\right]'=0\label{eq:vardpsidt}\\
\frac{\partial\bar{R}}{\partial (\partial_t h)}-\left[\frac{\partial\bar{R}}{\partial (\partial_t h')}\right]' &= \frac{\partial \fk}{\partial h} r h +\fk r -\alpha \cos\psi +\beta \sin\psi-\zeta'=0.\label{eq:vardhdt}
 \end{align}

To eliminate Lagrange multipliers, we take the derivative of Eq.~(\ref{eq:vardhdt}) with respect to $u$ and substitute $\alpha'$ as given in Eq.~(\ref{eq:alphaprime}) and $\beta'$ as given in Eq.~(\ref{eq:betaprime}). In addition, we substitute $\psi''$ as determined by Eqs.~(\ref{eq:vardpsidt}) and (\ref{eq:HelfAxi}) and use the geometric relations $r'=h\cos\psi$ and $z'=\sin\psi$, which finally leads to
 \begin{equation}\label{eq:zetadd}
     \zeta '' = 0.
 \end{equation}
 From the boundary term of the $h$ variation, $(\zeta\delta h)|_0^1=0$, and the fact that the map $h$ is not fixed before the variation, i.e. $\delta h(0),\delta h(1)\ne0$, we find boundary conditions $\zeta(0) = \zeta(1) = 0$. Consequently, Eq.~(\ref{eq:zetadd}) implies $\zeta = 0$ for all $u\in[0,1]$. Equations~(\ref{eq:vardpsidt}) and (\ref{eq:vardhdt}) therefore can be solved for the remaining Lagrange multipliers, which yields
 \begin{align}
\alpha &= \left[\bteun\sin\psi+(\bteuu-\gamma)\cos\psi\right]r \label{eq:alpha}\\
\beta &= \left[\bteun\cos\psi-(\bteuu-\gamma)\sin\psi\right]r. \label{eq:beta}
 \end{align}

Substituting $\beta$ given in Eq.~(\ref{eq:beta}) into Eq.~(\ref{eq:betaprime}) and rearranging terms yields
 \begin{align}
 \label{eq:betaprime2}
     &\frac{[\bteun]'}{h}+\frac{\cos\psi}{r}\bteun-\Cpp\bteuu-\tan\psi \left(\frac{[\bteuu]'}{h}+\bteun \Cuu - \frac{\gamma'}{h}\right)-\bteuu \Cuu\nonumber\\
     &= \left[\btduu\frac{\psi'} {h}+\btdpp\frac{\sin\psi}{r}  -  \fn  -  \frac{\partial \fk}{\partial C_0}\frac{C_0'}{h}\tan\psi\right].
 \end{align}
Using the the meridional component of the divergence of the equilibrium stress [see Eqs.~(\ref{eq:divTeu}) and (\ref{eq:divTu})], several terms in Eq.~(\ref{eq:betaprime2}) can be eliminated and we see by comparison with Eqs.~(\ref{eq:divTn}) that the remaining terms are together equivalent to the normal force balance
\begin{equation}
{\bf n} \cdot {\rm div}(\hat{\bf{T}}_d+{\bf T}_e)  = 0,     
\end{equation}
as required.

\section{Boundary conditions from variations}\label{app:boundary}
\subsection{Boundary conditions with Eulerian parameterization}
Note that the Rayleigh functional Eq.~(\ref{eq:rayleighaxi}) contains a non-trivial boundary term that generally modifies boundary terms of the variation. For the variations computed in Sec.~\ref{app:vareul} they read
\begin{align}
\left[\frac{\partial R}{\partial[(\bar{v}_u)']}+f_Hr\right]\delta \bar{v}_u\bigg\vert _{u=0}^{u=1} &=2\pi r\,\bar{\mathbf{e}}_u\cdot(f_H \mathbf{G}+\mathbf{T}_d)\cdot\bar{\mathbf{e}}_u\delta \bar{v}_u\bigg\vert_{u=0}^{u=1}\label{eq:bcvu}\\
\frac{\partial R}{\partial[ (\bar{v}_{\phi})']}\delta \bar{v}_{\phi}\bigg\vert _{u=0}^{u=1} &= 2\pi r\,\bar{\mathbf{e}}_u\cdot(f_H \mathbf{G}+\mathbf{T}_d)\cdot\bar{\mathbf{e}}_{\phi}\delta \bar{v}_{\phi}\bigg\vert_{u=0}^{u=1}\label{eq:bcvphi}\\
\left[\frac{\partial R}{\partial[ (v_n)']}-\frac{d}{du}\left(\frac{\partial R}{\partial[ (v_n)'']} \right)\right]\delta \bar{v}_n\bigg\vert _{u=0}^{u=1} &= 2\pi r\,\bar{\mathbf{e}}_u\cdot\mathbf{T}_e\cdot\mathbf{n} \delta v_n\bigg\vert_{u=0}^{u=1}\label{eq:bcvn}\\
\frac{\partial R}{\partial[ (v_n)'']}\delta \bar{v}_n'\bigg\vert _{u=0}^{u=1} &= \frac{2\pi r}{h}\,\bar{\mathbf{e}}_u\cdot\mathbf{M}_e\cdot\mathbf{n} \delta v_n'\bigg\vert_{u=0}^{u=1},\label{eq:bcdvn}
\end{align}
where $\mathbf{M}_e=m^{ij}_e\mathbf{e}_i\otimes\mathbf{e}_j$ with $m^{ij}_e$ given in Eq.~(\ref{eq:me}). 

\subsection{Boundary conditions with SLE parameterization and geometric constraints}
For the variations computed in Sec.~\ref{app:varsle} they now read
\begin{align}
\left[\frac{\partial\bar{R}}{\partial[(\bar{v}_u)']}+\frac{\partial (f_H q^u r h)} {\partial \bar{v}_u}\right]\delta \bar{v}_u\bigg\vert _{u=0}^{u=1}&=2\pi r\,\bar{\mathbf{e}}_u\cdot(f_H \mathbf{G}+\mathbf{T}_d)\cdot\bar{\mathbf{e}}_u\delta \bar{v}_u\bigg\vert_{u=0}^{u=1}\\
\frac{\partial\bar{R}}{\partial[ (\bar{v}_{\phi})']}\delta \bar{v}_{\phi}\bigg\vert _{u=0}^{u=1} &= 2\pi r\,\bar{\mathbf{e}}_u\cdot(f_H \mathbf{G}+\mathbf{T}_d)\cdot\bar{\mathbf{e}}_{\phi}\delta \bar{v}_{\phi}\bigg\vert_{u=0}^{u=1}\label{eq:bcvphi2}\\
\left[ \frac{\partial\bar{R}}{\partial[(\partial_t r)']}+\frac{\partial (f_H q^u r h)} {\partial  (\partial_t r)} \right]\delta (\partial_t r)\bigg\vert _{u=0}^{u=1} &= (\alpha-f_Hr\cos\psi) \delta (\partial_t r)\bigg\vert_{u=0}^{u=1}\\
\left[\frac{\partial\bar{R}}{\partial[(\partial_t z)']}+\frac{\partial (f_H q^u r h)} {\partial  (\partial_t z)}     \right]\delta (\partial_t z)\bigg\vert _{u=0}^{u=1} &=  (\beta+f_Hr\sin\psi) \delta (\partial_t z)\bigg\vert_{u=0}^{u=1}\\
\frac{\partial\bar{R}}{\partial[(\partial_t \psi)']}\delta (\partial_t \psi)\bigg\vert _{u=0}^{u=1} &=  2\pi r\,\bar{\mathbf{e}}_u\cdot\mathbf{M}_e\cdot\mathbf{\bar{e}}_\phi \delta (\partial_t \psi)\bigg\vert_{u=0}^{u=1}.
\end{align}

%%%%%%%%%%%%%%%%%%
%%%%%%%%%%%%%%%%%%
\begin{figure*}[t]
	\includegraphics[width = 1\columnwidth]{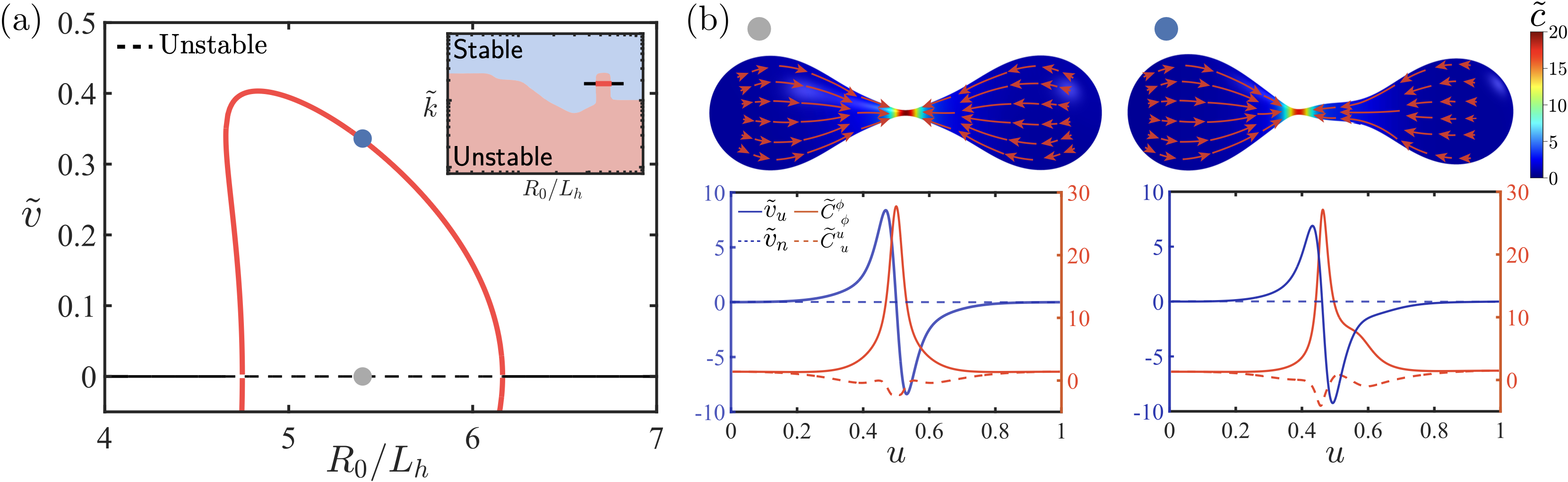}\vspace{-0.2cm}
	\caption{Symmetry-breaking and -restoring reentrant transition of stationary active fluid surfaces (see red line in Fig.~\ref{fig:furrowshsp}c, main text). \textbf{(a)}~In a specific region of the parameter space the symmetrically ingressed surface (translational velocity $\tilde{v}=0$) becomes unstable (dashed line) when increasing the inverse hydrodynamic length $R_0/L_h$ and two stable branches with asymmetrically places necks emerge (red solid lines). For orientation, the inset reproduces minimal features of Fig.~\ref{fig:furrowshsp}c in the main text. \textbf{(b)}~Degenerate stationary surface solutions that are symmetrically (unstable) and asymmetrically (stable) ingressed (see corresponding dots in (a)).}
	\label{fig:S1}
\end{figure*}
%%%%%%%%%%%%%%%%%%
%%%%%%%%%%%%%%%%%%

%%%%%%%%%%%%%%%%%%
%%%%%%%%%%%%%%%%%%
\begin{figure*}[t]
	\includegraphics[width = 0.8\columnwidth]{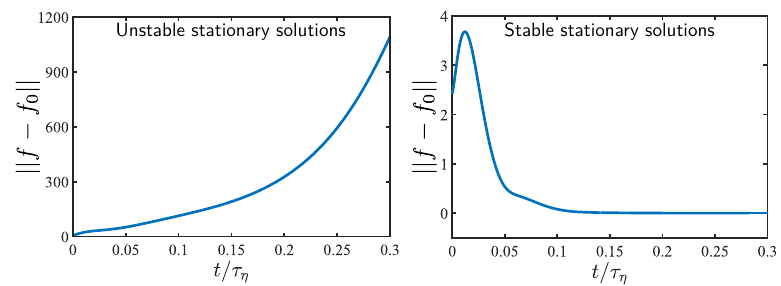}\vspace{-0.2cm}
	\caption{Stability testing of nonlinear stationary surface solutions. To test the stability of a numerically exact stationary solutions, say $f_0$, found by the approach described in Sec.~\ref{sec:dcomp}, we observe their response to small perturbations in dynamic simulations. As measure for the deviation of these simulations, we consider $||f-f_0||:=||r(u,t)-r_0(u)||_2+||z(u,t)-r_0(u)||_2+||c(u,t)-c_0(u)||_2$, where $||\cdot||_2$ denotes the $L_2$ norm on $u\in[0,1]$. This test typically leads to an unambiguous diverging (left) or converging dynamics (right), from which we conclude the corresponding point of the branch to be unstable or stable, respectively. Curves show exemplary responses from the unstable and stable parameter regions in Fig.~\ref{fig:furrowshsp}c.}
	\label{fig:stabtest}
\end{figure*}
%%%%%%%%%%%%%%%%%%
%%%%%%%%%%%%%%%%%%

\end{document}